\pdfoutput=1
\documentclass{aa}  
\usepackage{graphicx}
\usepackage{txfonts}
\usepackage{xcolor}  
\usepackage{multirow} 
\usepackage{amsmath} 
\usepackage{hyperref}
\begin{document}

   \title{Multi-wavelength observations of the Luminous Fast Blue Optical Transient AT\,2023fhn}

   \subtitle{Up to $\sim$200 days post-explosion}

   \author{A. A. Chrimes\inst{1}\fnmsep\inst{2}\thanks{ESA Research Fellow}, 
          D. L. Coppejans\inst{3},
          P. G. Jonker\inst{2},
          A. J. Levan\inst{2}\fnmsep\inst{3},
          P. J. Groot\inst{2}\fnmsep\inst{4}\fnmsep\inst{5},
          A. Mummery\inst{6}
          \and
          E. R. Stanway\inst{3}
          }

   \institute{European Space Agency (ESA), European Space Research and Technology Centre (ESTEC), Keplerlaan 1, 2201 AZ Noordwijk, the Netherlands \\
              \email{ashley.chrimes@esa.int}
         \and
            Department of Astrophysics/IMAPP, Radboud University, PO Box 9010, 6500 GL Nijmegen, The Netherlands
         \and
            Department of Physics, University of Warwick, Gibbet Hill Road, CV4 7AL Coventry, United Kingdom 
         \and
            Inter-University Institute for Data Intensive Astronomy, Department of Astronomy, University of Cape Town, Private Bag X3, Rondebosch 7701, South Africa
         \and
            South African Astronomical Observatory, P.O. Box 9, 7935 Observatory, South Africa
         \and
            Oxford Astrophysics, Denys Wilkinson Building, Keble Road, Oxford, OX1 3RH, United Kingdom
             }

   \date{Received September 15, 1996; accepted March 16, 1997}

 
  \abstract
   {Luminous Fast Blue Optical Transients (LFBOTs) are a class of extragalactic transients notable for their rapid rise and fade times, blue colour and accompanying luminous X-ray and radio emission. Only a handful have been studied in detail since the prototypical example AT\,2018cow. Their origins are currently unknown, but ongoing observations of previous and new events are placing ever stronger constraints on their progenitors.}
   {We aim to put further constraints on the LFBOT AT\,2023fhn, and LFBOTs as a class, using information from the multi-wavelength transient light-curve, its host galaxy and local environment.}
   {Our primary results are obtained by fitting galaxy models to the spectral energy distribution of AT\,2023fhn's host and local environment, and by modelling the radio light-curve of AT\,2023fhn as due to synchrotron self-absorbed emission from an expanding blast-wave in the circumstellar medium.}
   {We find that neither the host galaxy nor circumstellar environment of AT\,2023fhn are unusual compared with previous LFBOTs, but that AT\,2023fhn has a much lower X-ray to ultraviolet luminosity ratio than previous events.}
   {We argue that the variety in ultraviolet-optical to X-ray luminosity ratios among LFBOTs is likely due to viewing angle differences, and that the diffuse, yet young local environment of AT\,2023fhn - combined with a similar circumstellar medium to previous events - favours a progenitor system containing a massive star with strong winds. Plausible progenitor models in this interpretation therefore include black hole/Wolf-Rayet mergers or failed supernovae.}

               
   \titlerunning{Multi-wavelength observations of AT\,2023fhn}
   \authorrunning{A. A. Chrimes et al.}
   
   \maketitle
%


\section{Introduction}
Luminous Fast Blue Optical Transients (LFBOTs) are a rare class of rapidly-evolving, hot, multi-wavelength extragalactic transients. The prototypical example, AT\,2018cow \citep[`the Cow',][]{2018ApJ...865L...3P}, is the nearest and best-studied event of this class so far. Its characteristic early-time features include a peak optical absolute magnitude of $\sim$--20 with a rapid rise and decay timescale of $\sim$5 days, constraining the $^{56}$Ni mass in the ejecta to $<$$0.004$\,M$_{\odot}$, and ruling out standard core-collapse or thermonuclear supernova models \citep{2019MNRAS.484.1031P,2019ApJ...872...18M}. The optical-ultraviolet (UV) emission is well fit by a hot black-body and power-law component, where the black-body temperature was $\sim 3\times10^4$\,K initially, falling to $\sim 1.5\times 10^4$\,K over two weeks \citep{2018ApJ...865L...3P,2019MNRAS.484.1031P}. The optical spectra were largely featureless, with broad hydrogen absorption features (indicative of a high outflow velocity) appearing and disappearing between 2--8 days, and narrow He lines appearing after $\sim$20 days. At other wavelengths, AT\,2018cow was X-ray and radio bright \citep{2018MNRAS.480L.146R,2019ApJ...872...18M,2019ApJ...871...73H,2021ApJ...912L...9N}. The X-ray emission was well in excess of power-law extrapolations from the radio \citep[e.g.][]{2019ApJ...871...73H} and was also highly variable after a break in the light-curve, which declined as $L \propto t^{-2}$ after $\sim$20 days \citep{2024ApJ...963L..24M}. The broadband X-ray spectrum, and X-ray variability, cannot be explained by an external shock origin, and the synchrotron self-absorbed radio emission - consisting of a slow rise and rapid decay - was instead smoothly evolving, indicating a distinct physical origin from the highly variable X-rays. An interpretation is that a central engine powers the X-ray emission, while an expanding blast wave produces the radio emission \citep[e.g.][]{2019ApJ...871...73H,2019ApJ...872...18M}. The slow radio variability timescale set the size of the emission region at 5--6 days at $<3\times10^{15}$\,cm, while the X-ray variability gave a length scale $\sim$five times smaller \citep{2019ApJ...871...73H}. Therefore, the X-rays appear to originate from a central engine or internal shock, while the radio emission is generated externally. A claim of quasi-periodic oscillations in the X-rays can interpreted as evidence for a $<850 M_{\odot}$ central engine \citep{2022NatAs...6..249P}, while a separate claim of $\sim$250s quasi-periodicity instead implies an intermediate mass (10$^{3}$-10$^{5}$\,M$_{\odot}$) black hole \citep{2022RAA....22l5016Z}. Synchrotron modelling of the sub-millimetre and radio data revealed a mildly-relativistic expansion velocity ($\sim$0.1$c$) into a wind-like extended circumstellar medium (CSM) with a high density of $\sim$10$^{5}$\,cm$^{-3}$ \citep{2019ApJ...872...18M,2019ApJ...871...73H}. For Wolf-Rayet-like wind speeds of $\sim$1000\,km\,s$^{-1}$, this implies a mass-loss rate $\dot{M} = 10^{-4}$-$10^{-3}$\, $M_{\odot}$\,yr$^{-1}$ \citep{2019ApJ...872...18M}. 

Since AT\,2018cow, several more LFBOTs have been discovered. Confirmed events include AT\,2018lug/ZTF\,18abvkwla \citep[`the Koala',][]{2020ApJ...895...49H}, CSS161010 \citep{2020ApJ...895L..23C}, AT\,2020xnd/ZTF\,20\,acigmel \citep[`the Camel',][]{2021MNRAS.508.5138P,2022ApJ...926..112B,2022ApJ...932..116H}, AT\,2020mrf \citep{2022ApJ...934..104Y}, AT\,2022tsd \citep[`the Tasmanian Devil',][]{2023RNAAS...7..126M} and AT\,2023fhn \citep[`the Finch',][]{2024MNRAS.527L..47C}. Despite variety (e.g. in peak luminosity), they share the same key features of hot, largely featureless spectra at early times, optical luminosities rivalling superluminous supernovae, plus bright X-ray and radio emission. They are estimated to occur at $<$0.1\% of the local core-collapse supernova rate \citep{2023ApJ...949..120H}. 

Recent developments have provided further insight into the origin of LFBOTs. Polarimetry of AT\,2018cow demonstrated the emission region to be highly aspherical, indicative of an accretion disc \citep{2023MNRAS.521.3323M}. Unexpectedly, AT\,2018cow was found to be UV \citep{2022MNRAS.512L..66S,2023ApJ...955...43C,2023MNRAS.519.3785S,2023MNRAS.525.4042I} and X-ray \citep{2024ApJ...963L..24M} bright at late times, several years post-explosion. This emission has been interpreted as from a black hole accretion disc. Estimates for the black hole mass range from $\sim$10--100\,M$_{\odot}$ (super-Eddington accretion) to $\sim$10$^{3}$--10$^{4}$\,M$_{\odot}$ \citep[sub-Eddington, from X-ray observations,][]{2024ApJ...963L..24M} and $\sim$1000\,M$_{\odot}$ \citep[UV observations,][]{2023MNRAS.525.4042I}. Magnetar central engine models struggle to produce both the early and late UV emission \citep{2023ApJ...955...43C}. Further evidence for a black hole accretion scenario comes from minute-long optical flares, up to several months post-explosion, from AT\,2022tsd \citep{2023Natur.623..927H}. An interpretation is that the central engine is undergoing highly variable, short-lived bursts of accretion.

Several models have been put forward to explain LFBOTs. Tidal disruptions of compact, hydrogen-poor stars (such as white dwarfs) around intermediate mass black holes (IMBHs) can plausibly explain the optical rise and fall timescale, spectral features and X-ray variability timescale \citep{2019MNRAS.484.1031P,2019MNRAS.487.2505K}. However, the dense CSM inferred from radio observations is hard to explain in such a scenario \citep[e.g.][]{2019ApJ...872...18M}. Other possibilities include failed supernovae, in which a black hole is formed and the emission is powered by accretion onto the natal black hole rather than radioactive decay in the ejecta \citep{2019MNRAS.484.1031P,2019MNRAS.485L..83Q}, choked jets \citep[e.g.][]{2022MNRAS.513.3810G,2022RAA....22e5010S}, highly aspherical supernovae \citep[`ellipsars',][]{2022ApJ...931L..16D}, and the mergers of compact objects and/or massive stars \citep{2019MNRAS.487.5618L,2020ApJ...897..156U,2020ApJ...892...13S}, such as black holes and Wolf-Rayet stars \citep{2022ApJ...932...84M}. A dense outflow from the progenitor may result in dust echoes \citep{2023ApJ...944...74M}. CSM shock interaction models have also been put forward \citep[e.g.][]{2019MNRAS.488.3772F,2021ApJ...910...42X,2022ApJ...926..125P,2023arXiv230403360K}, but the X-ray variability, broadband spectral evolution, late-time UV/X-ray emission from AT\,2018cow and giant optical flares from AT\,2022tsd all indicate the presence of a central engine.

In this paper, we present multi-wavelength radio, optical, UV, and X-ray observations of the LFBOT AT\,2023fhn up to $\sim$200 days post-explosion. We place AT\,2023fhn in the context of other LFBOTs so far, in terms of its host galaxy, optical/UV/X-ray light-curve, and radio emission, with the event energetics and blast wave properties inferred from synchrotron modelling of the radio observations. Throughout, we use a flat $\Lambda$CDM cosmology with $\Omega_{\rm m}$=0.3 and $H_{0}$=70\,kms$^{-1}$Mpc$^{-1}$. All magnitudes are reported in the AB system \citep{1982PASP...94..586O}.

\section{Observations and data reduction}~
\subsection{X-ray}
We obtained four epochs of Chandra X-ray Observatory (CXO) ACIS-S observations of AT\,2023fhn up to $\sim$200 days post-explosion. The epochs consist of 1, 2, 6 and 14 observations, respectively (full details are provided in Table \ref{tab:xraydata}). The data are reduced, and transient fluxes measured, with standard {\sc CIAO} \citep[v4.13, caldb v4.9.3,][]{2006SPIE.6270E..1VF} procedures. The images are reprocessed and filtered to the energy range 0.5-7.0\,keV. {\sc wavdetect} is used to find point sources, and {\sc srcflux} used to measure the flux (or upper limits) at the location of AT\,2023fhn. We merged the datasets in each of the four epochs (with {\sc merge$\_$obs}) to increase the signal-to-noise ratio. The mean (mid-point, exposure-time weighted) observation times of these epochs are 15.0, 28.9, 64.5 and 210.9 days (since JD--2460045, or 12:00 UT on 10-Apr-2023). The total exposure times per epoch are $\sim$30, 60,  83 and 193\,ks respectively. Finally, the fluxes are de-absorbed by assuming a photon index $\Gamma=2$ \citep[e.g.][]{2018MNRAS.480L.146R}, and a Galactic neutral hydrogen column density of $N_{\rm H}=2.78\times10^{20}$\,cm$^{-2}$ \citep{1990ARA&A..28..215D}. 

\begin{table}
\centering 
\caption{All CXO observations of AT\,2023fhn from programme 24500143 (PI: Chrimes). ObsID, exposure start times (since JD-2460045) and data mode are listed. All observations are made with ACIS-S. The fluxes F$_{\rm X}$ are unabsorbed and measured in the energy range 0.5-7.0\,keV. Individual observations in each of the four epochs are merged as indicated. Uncertainties are given at 1$\sigma$, upper limits at 2$\sigma$.}
\label{tab:xraydata}
\begin{tabular}{ccccl}
\hline %
\hline %
ObsID & Start date & t$_{\rm exp}$ & Data mode & F$_{\rm X}$ \\
 & JD-2460045 & ks & & erg\,s$^{-1}$\,cm$^{-2}$ \\
\hline %
26624	&	14.78957	&	29.68	&	FAINT	& (7.6$^{+2.2}_{-1.8}$)$\times10^{-15}$ \\
26625	&	27.98310	&	29.68	&	FAINT	& \hspace{-0.7cm}\multirow{2}{*}{$\left.\begin{array}{l}
                  \\
                \end{array}\right\rbrace \hspace{0.1cm} $(4.5$^{+4.7}_{-2.9}$)$\times10^{-16}$ } \\
27833	&	29.47145	&	29.67	&	FAINT &	\\
26626	&	61.80356	&	16.88	&	VFAINT & \hspace{-0.7cm}\multirow{6}{*}{$\left.\begin{array}{l}
                  \\
                  \\
                  \\
                  \\
                  \\
                \end{array}\right\rbrace \hspace{0.1cm} <8.2\times10^{-16}$ } \\
27895	&	62.33516	&	10.94	&	VFAINT &	\\
27835	&	65.12251	&	13.89	&	FAINT &	\\
27905	&	65.45429	&	13.89	&	FAINT &	\\
27906	&	65.79704	&	13.89	&	FAINT &	\\
27907	&	66.13969	&	13.89	&	FAINT &	\\
26627	&	198.66317	&	10.74	&	VFAINT & \hspace{-0.7cm}\multirow{14}{*}{$\left.\begin{array}{l}
                  \\
                  \\
                  \\
                  \\
                  \\
                  \\
                  \\
                  \\
                  \\
                  \\
                  \\
                  \\
                  \\
                \end{array}\right\rbrace \hspace{0.1cm} <3.5\times10^{-16}$ } \\
28997	&	198.96634	&	10.74	&	VFAINT &	\\
28998	&	199.27127	&	11.12	&	VFAINT &	\\
27837	&	205.89744	&	13.4	&	VFAINT &	\\
29031	&	206.23509	&	13.3	&	VFAINT &	\\
29032	&	206.57170	&	13.5	&	VFAINT &	\\
29034	&	207.96147	&	10.93	&	VFAINT &	\\
29033	&	208.26639	&	14.39	&	VFAINT &	\\
27838	&	215.80840	&	16.85	&	VFAINT &	\\
29054	&	216.19957	&	17.84	&	VFAINT &	\\
29056	&	216.60163	&	14.39	&	VFAINT &	\\
28991	&	218.76822	&	9.94	&	VFAINT &	\\
28999	&	219.12318	&	18.69	&	VFAINT &	\\
29055	&	219.54352	&	16.85	&	VFAINT &	\\
\hline
\end{tabular}
\end{table}

\subsection{UV-optical}
\begin{figure}
   \centering
   \includegraphics[width=\columnwidth]{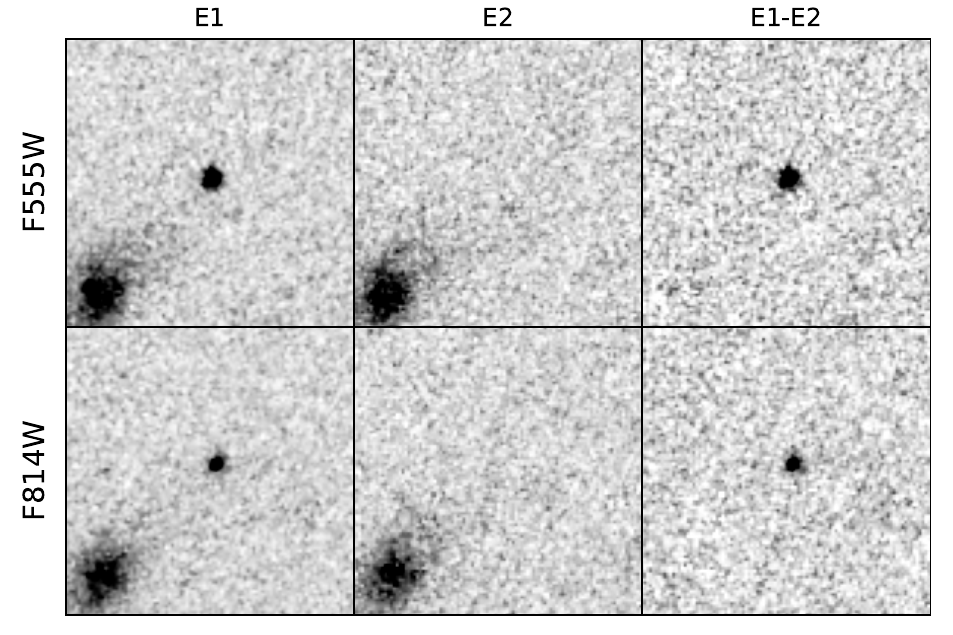}
      \caption{F555W and F814W HST/WFC3 image stamps at the location of AT\,2023fhn in the first epoch (May 2023, left) and the second epoch (October 2023, centre). The right-hand panels show difference images (epoch 1 - epoch 2). North is up, East is left, and the stamps are 2.5\,arcsec on each side.}
         \label{fig:hst}
   \end{figure}

A second epoch of HST imaging was obtained on 23/24 October 2023 \citep[the first was on 17 May 2023,][]{2024MNRAS.527L..47C}, using the WCF3 instrument and six filters ($F225W, F336W, F555W, F763M, F814W, F845M$). Full details are given in Table \ref{tab:hstdata}. The data are reduced with {\sc drizzlepac} \citep{2021AAS...23821602H}, re-drizzling the charge-transfer-efficiency-corrected $\_${\sc flc} input images with North oriented up and a final pixel scale of 0.025\,arcsec\,pixel$^{-1}$ ({\sc pixfrac}=0.8). Image stamps around the location of AT\,2023fhn in epochs 1 and 2 are shown in Figure \ref{fig:hst}. Visible in the bottom left is the presumed satellite of the larger spiral to the south (see Figure \ref{fig:hostSEDs}). Both galaxies lie at a common redshift of $\sim$0.24 \citep{2023TNSAN..93....1H,2024MNRAS.527L..47C}.

\begin{table}
\centering 
\caption{All HST data for AT\,2023fhn, from program 17238 (PI: Chrimes). Filter, exposure start times (JD-, where is 12:00 UT on 10-APR-2023) and exposure durations t$_{\rm exp}$ are given. All observations are with WFC3 in the UVIS channel.}
\label{tab:hstdata}
\begin{tabular}{ccc}
\hline %
\hline %
Filter & Start date & t$_{\rm exp}$ \\
 & JD--2460045 & s \\
\hline %
F555W &	36.87666    & 1092	\\
F814W &	36.89272    & 1092	\\
F555W &	196.42824    & 990	\\
F814W &	196.44313    & 1092	\\
F225W &	196.49431    & 1068	\\
F336W &	196.51027    & 1068	\\
F845M &	196.56034   & 990	\\
F763M &	196.57515    & 1068	\\
\hline
\end{tabular}
\end{table}
   
\subsection{Radio}
We obtained radio observations with the Karl G. Jansky Very Large Array (VLA) between 22 Apr 2023 and 16 December 2024 (programme SC240143, PI: Chrimes). Details of the observations are listed in Table \ref{tab:radio_table}. The observations were taken in standard phase-referencing mode using 3C286 as a flux density and bandpass calibrator, with ICRF J101447.0+230116, FIRST J101644.3+203747, FIRST J101353.4+244916 and ICRF J095649.8+251516 as complex gain calibrators. The observations were calibrated using the VLA Calibration Pipeline versions 2023.1.0.124 and 2022.2.0.64 in \textsc{CASA} 6.5.4 and 6.4.1 respectively, with additional manual flagging. The images were created using the tclean task in \textsc{CASA} with Briggs weighting with a robust parameter of 1. In the observations where the source was not detected we quote the upper limit on the flux density as three times the local RMS. The one exception to this is during the last epoch (see Table \ref{tab:radio_table}) where the synthesized beam (resolution element) was large and included other sources. In this case we quoted the upper limit as the flux density at the source location. For the observations where we detected the target, we fitted the flux density using the imfit task within \textsc{CASA} and constrained the fit to the synthesized beam.   

The observations up to $\sim$12 days post JD-2460045 are already published \citep{2024MNRAS.527L..47C} and all produced non-detections. In the $\sim$87--95 day and $\sim138$ day epochs we have sufficient data points for fitting a synchrotron self-absorbed spectrum. The K$_{\rm U}$ band (15\,GHz) data point at 138 days has sufficient signal-to-noise to split into 3 (centred on 13, 15 and 17\,GHz), as listed in Table \ref{tab:radio_table}, increasing the points at $\sim$138 days to 7 (with 6 detections). We fit a self-absorbed synchrotron model to the $\sim$87-95 and $\sim$138 day epochs in Section \ref{sec:radio_results}.

\begin{table}
	\centering
	\caption{AT\,2023fhn flux densities from our VLA programme (SC240143, PI: Chrimes). Observation start times are listed with respect to JD--2460045 (12:00 on 10-Apr-2023). The quoted uncertainties do not include the systematic uncertainty of 5\% on the absolute flux calibration at these frequencies. Upper limits are given as 3 times the local RMS. $^a$The resolution of the last three observations was lower (as the VLA was in D configuration at the time), so we could not remove contaminating sources at the target location and have consequently listed the upper limit as the flux density at the source location.}
	\label{tab:radio_table}
	\begin{tabular}{ccccc} 
		\hline
		\hline
		  Start date & Freq. & Bandwidth & t$_{\rm exp}$ & Flux Density \\
            JD--2460045 & GHz & GHz & min. & $\mu\rm{Jy/beam}$\\
		\hline
            11.80740 & 1.52 & 1.024 & 38 & $<$130 \\
            11.78309 & 3.00 & 2.048 & 32 & $<$35 \\
            11.76507 & 6.00 & 4.096 & 23 & $<$18 \\
            11.74688 & 10.00 & 4.096 & 23 & $<$18 \\
            11.72090 & 15.08 & 6.144 & 35 & $<$11 \\
            11.69229 & 22.00 & 8.192 & 35 & $<$17 \\
            11.66552 & 33.00 & 8.192 & 33 & $<$25 \\
            87.59185 & 1.52 & 1.024 & 39 & $<$45\\
            87.56657 & 3.00 & 2.048 & 33 & 110$\pm$8\\
            87.54257 & 6.00 & 4.096 & 32 & 128$\pm$5\\
            95.58690 & 10.00 & 4.096 & 31 & 105$\pm$7 \\
            95.56438 & 15.00 & 6.144 & 30 & 71$\pm$7 \\
            95.52738 & 22.00 & 8.192 & 47 & 60$\pm$10 \\
            137.17440 & 1.52 & 1.024 & 39 & $<$58	 \\
            137.14178 & 3.00 & 2.048 & 44 & 110$\pm$10 \\
            137.10567 & 6.00 & 4.096 & 50 & 221$\pm$5 \\
            138.16972 & 10.00 & 4.096 & 42 & 197$\pm$6 \\
            138.13664 & 13.00 & 2.050 & 42 & 180$\pm$10 \\
            138.13664 & 15.00 & 2.050 & 42 & 160$\pm$10 \\
            138.13664 & 16.96 & 2.050 & 42 & 140$\pm$20 \\
            249.95139 & 6.00 & 4.096 & 35 & $<$311$^a$ \\
            249.90997 & 10.00 & 4.096 & 57 & $<$181$^a$ \\
            249.86861 & 15.08 & 6.144 & 57 & $<$125$^a$ \\
		\hline
	\end{tabular}
\end{table}

\section{Environmental analysis}~
\subsection{Local environment}\label{sec:local}
The second epoch of HST imaging presented in this paper allows us to examine the environment directly underlying the transient after it has faded. As noted by \citet{2024MNRAS.527L..47C}, there is diffuse emission in the vicinity of the transient. To characterise this faint underlying population, we place 0.2\,arcsec (and 0.4\,arcsec) apertures at the location of AT\,2023fhn in all six epoch 2 images. The images are aligned with $x$-$y$ shifts using 5 common point sources in every image, with respect to the location of AT\,2023fhn in the epoch 1 F555W image. The rms of these relative astrometric alignments is $\sim$5--10\,mas, better than the absolute astrometry of the images (which have been aligned with the {\it Gaia} DR3 reference frame), and much smaller than the aperture size. We perform photometry with {\sc photutils}, estimating the background with either the median image background (with {\sc medianbackground}) or an annulus (1.5 to 4 times the aperture radius, with pixels values clipped at 3$\sigma$). The appropriate encircled energy corrections for each filter and aperture are applied. Magnitudes are then calculated using the {\sc photplam} and {\sc photflam} header keywords\footnote{\url{https://hst-docs.stsci.edu/wfc3dhb/chapter-9-wfc3-data-analysis/9-1-photometry}}, and are listed in Table \ref{tab:phot_table}. The only detections are in F555W and F814W. To investigate the nature of these detections, we place eight 0.4\,arcsec apertures at equal spacing around the location of AT\,2023fhn in a circle of radius 20 pixels (0.5\,arcsec). With the F555W filter and median background subtraction, we have significant detections in 5/8 apertures, with a mean magnitude of 25.9$\pm$0.6 in these apertures - consistent with the measurement at the precise location of AT\,2023fhn. This demonstrates that the emission in this area is from an extended, diffuse background, rather than any significant contribution by residual light from AT\,2023fhn. This can also be seen in Table \ref{tab:phot_table}, where the magnitudes calculated with annulus background subtraction are fainter, since the local background is elevated. Larger apertures also give brighter magnitudes, despite encircled energy correction (unlike point sources in the field). For this reason, we exclusively use median background subtracted measurements - which are not biased by the nearest pixels in the image - in the following analysis. 

We disfavour a significant contribution from a compact cluster at this specific location, which would appear as a point source in the image given the physical scale at this redshift of $\sim$100\,pc\,pixel$^{-1}$. However, the presence of a globular cluster \citep[which would favour an IMBH interpretation, e.g.,][]{2013A&A...555A..26L} cannot be ruled out, as even the brightest globular clusters would be far below detection limits at this distance and limiting magnitude \citep{2024MNRAS.527L..47C}. Shifting the circle of apertures 5\,arcsec to the north, well away from the galaxies, we find non-detections in all eight apertures with a 3$\sigma$ upper limit of 26.7. We therefore conclude that there is extended, diffuse emission from an underlying stellar population at the location of AT\,2023fhn.

We now estimate the age and dust extinction of this underlying population. First, we correct for the (low) Galactic extinction of $E(B-V)$=0.0254 \citep{2011ApJ...737..103S}\footnote{\url{https://irsa.ipac.caltech.edu/applications/DUST/}} using the filter effective wavelengths \citep[][]{2012ivoa.rept.1015R,2020sea..confE.182R} and the Python {\sc extinction} package \citep{2016zndo....804967B} with a \citet{1999PASP..111...63F} extinction law and $R_{\rm V}=3.1$. To estimate the age and local (intrinsic) extinction, we fit the Galactic-extinction corrected $F225W, F336W, F555W$ and $F814W$ photometry (both 0.2 and 0.4\,arcsec apertures, median background subtracted) to BPASS \citep[Binary Population and Spectral Synthesis v2.1,][]{2017PASA...34...58E,2018MNRAS.479...75S} single-age spectral templates. These are constructed by assuming that a stellar population of 10$^{6}$M$_{\odot}$ is formed instantaneously, and left to evolve with no further star formation. We use these simple stellar populations since the limited data available to model solely the local environment of AT\,2023fhn precludes a more complex procedure, including, for example, the star-formation history (however, see the next section). A fixed metallicity of half-Solar is adopted ($Z=0.01$ by mass fraction). We therefore simply fit for the age of the population, the luminosity (i.e. mass) of the stellar population is then allowed to freely vary to minimise $\chi^{2}$. Four data points are used ($F225W, F336W, F555W$ and $F814W$). Formal flux measurements are used for $F225W$ and $F336W$; these produce magnitudes of 29.0$^{+1.7}_{-2.0}$ and 27.3$^{+0.6}_{-1.4}$ respectively (with a 0.2\,arcsec aperture) and 26.9$^{+0.7}_{-1.0}$ and 26.7$^{+0.6}_{-1.8}$ (with a 0.4\,arcsec aperture). We therefore have 2 fit parameters and 4 data points for 2 degrees of freedom. Fitting is performed by multiplying the (de-redshifted) filter response curves \citep[][]{2012ivoa.rept.1015R,2020sea..confE.182R} with the BPASS spectra to extract fluxes and hence magnitudes from the spectra. These are compared with the absolute magnitudes in each filter, after correction for a range of intrinsic extinction values from $A_{\rm V}$=0.0 to 4.0. The intrinsic extinction correction uses the rest-frame effective wavelength of each filter. The $F763M$ and $F845M$ filters are not used in this fit since the upper limits are shallower than the $F555W$ and $F814W$ detections, and so provide no additional constraints.

\begin{figure*}
   \centering
   \includegraphics[width=\columnwidth]{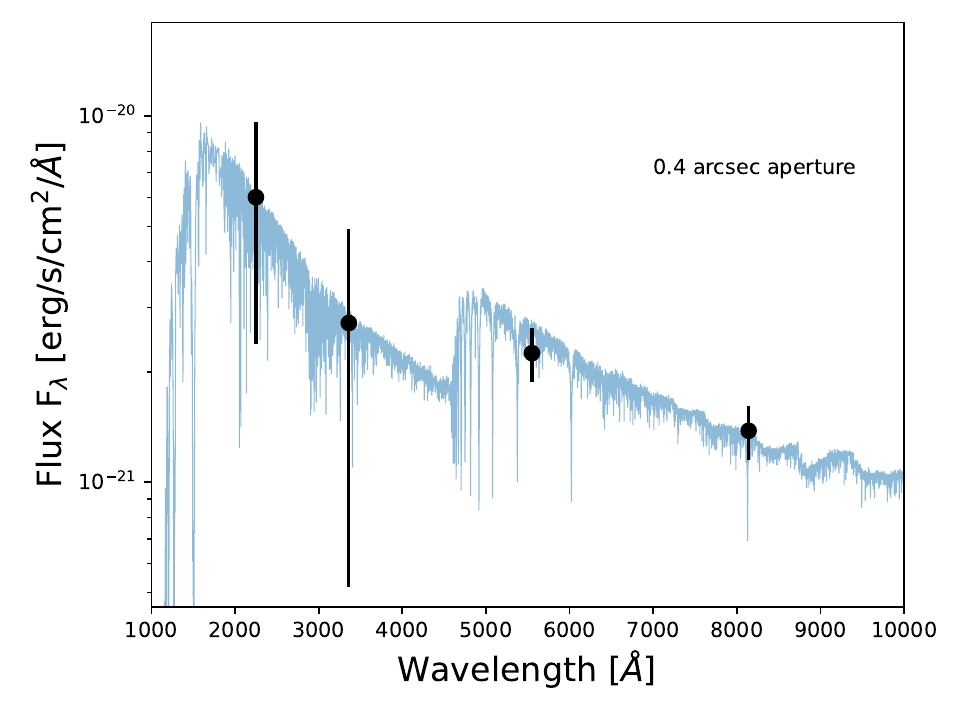}
   \includegraphics[width=\columnwidth]{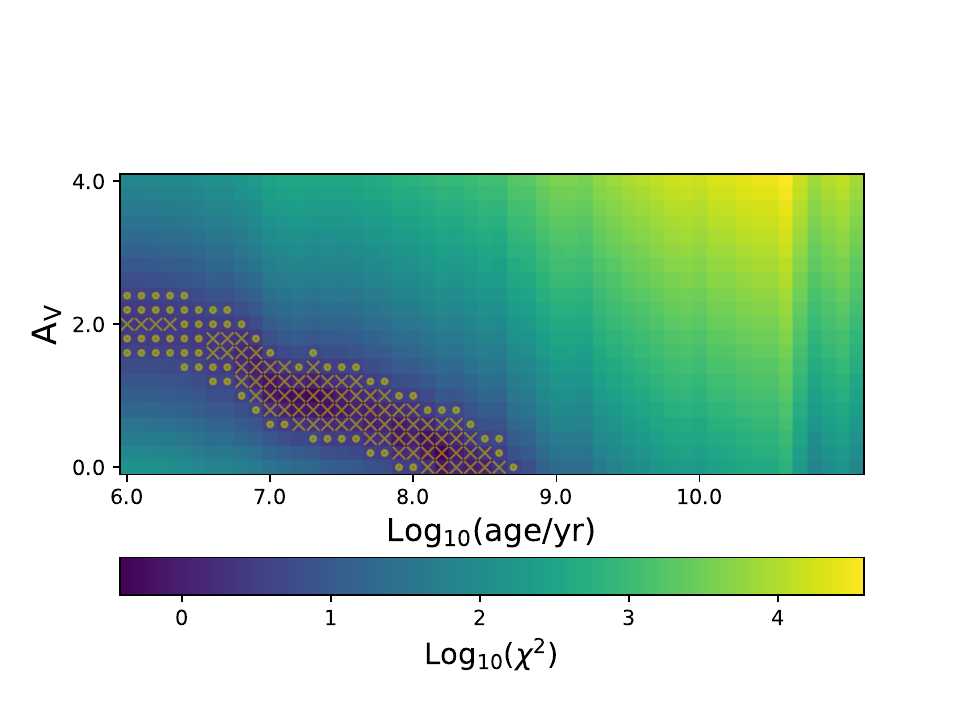}
   \includegraphics[width=\columnwidth]{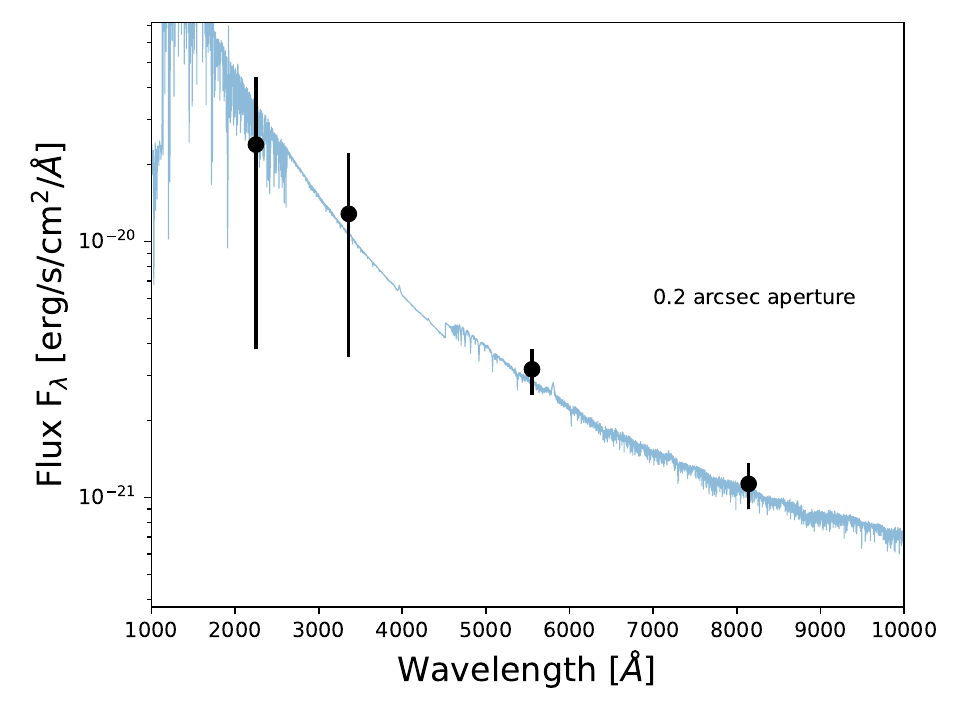}
   \includegraphics[width=\columnwidth]{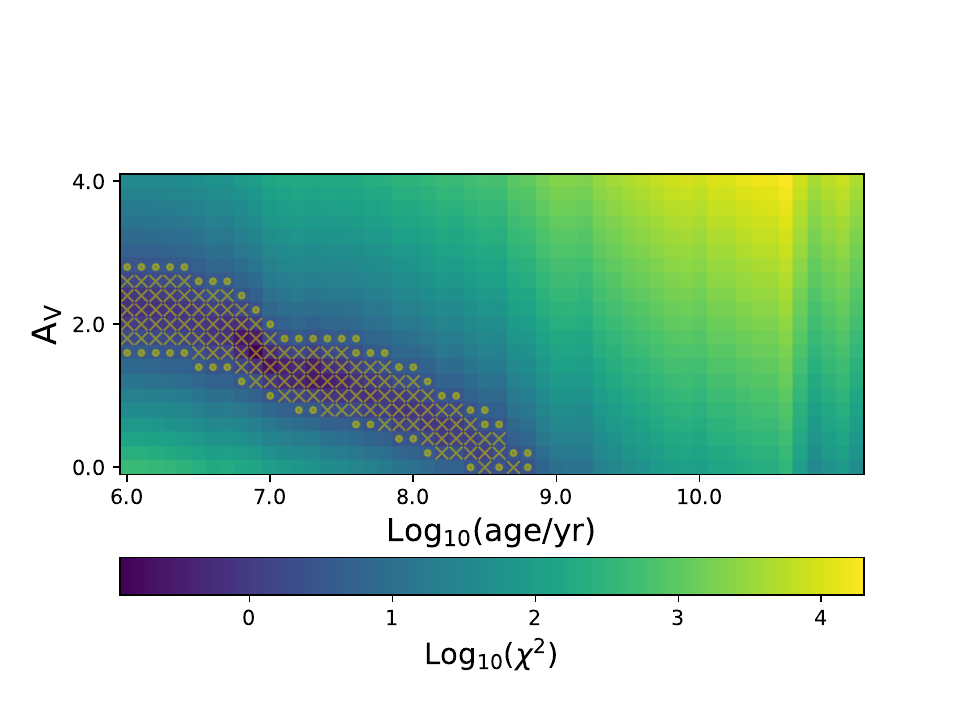}
      \caption{Epoch 2 {\em HST} photometry at the location of AT\,2023fhn, corrected for Galactic extinction, and fit to BPASS single age spectral models while allowing the (intrinsic) extinction to vary. Top row, left: the best fit single-age BPASS spectrum. Wavelengths are observer frame. The photometry was measured with median background subtraction and a 0.4\,arcsec aperture, and is corrected for both Galactic and instrinsic extinction. Top row, right: the colourmap corresponds to the fit log$_{10}$($\chi^{2}$) as a function of intrinsic extinction and age. We find a best fitting combination of Log(age/yr)=$8.2^{+0.4}_{-2.2}$ and A$_{\rm V}=0.2^{+1.8}_{-0.2}$ \citep[uncertainties at 68\% confidence,][]{1976ApJ...210..642A}. Pixels in the parameter space grid within the 68\% and 90\% confidence regions are indicated by crosses (within 68\%) and dots (in the 68--90\% range). Bottom row: as for the upper row, but using a 0.2\,arcsec aperture for photometry, with which we find Log(age/yr)=$6.9^{+1.8}_{-0.9}$ and A$_{\rm V}=1.6^{+1.0}_{-1.6}$.}
         \label{fig:diffuse_age}
   \end{figure*}

The results are shown in Figure \ref{fig:diffuse_age}. The left-hand panels show the best-fit single-age BPASS spectra. The right-hand panels show log$_{10}(\chi^{2})$ across the parameter space. Each pixel represents a unique combination of $A_{\rm V}$ and a BPASS simple stellar population at a given age. The 68\% and 90\% confidence intervals are indicated by crosses and dots respectively \cite[where the $\Delta \chi^{2}$ intervals are from][]{1976ApJ...210..642A}. We can see that the choice of a 0.2 or 0.4\,arcsec aperture makes little difference, likely because the increased aperture size simply captures more of the same diffuse background light at that position, without changing the colours significantly. In both cases, there is a degeneracy between age and extinction, and while the uncertainties are large, population ages in excess of a few hundred Myr are disfavoured. The range of possible ages for this diffuse emission is therefore broadly consistent with the core-collapse supernova delay time distribution \citep[e.g.][]{2019MNRAS.482..870E}.

We also measure the local surface brightness in epoch 2 (in a 0.5 arcsec radius around AT\,2023fhn's position), giving 25.1\,mag\,arcsec$^{-2}$ in $F555W$ and 24.65\,mag\,arcsec$^{-2}$ in $F814W$. This compares well with the 25.2\,mag\,arcsec$^{-2}$ and 24.6\,mag\,arcsec$^{-2}$ values from the transient-subtracted images in Epoch 1 \citep[see][]{2024MNRAS.527L..47C}. The $F336W$ surface brightness is 25.76\,mag\,arcsec$^{-2}$, which after Galactic extinction correction is 25.27\,mag\,arcsec$^{-2}$. The rest-frame central wavelength of $F336W$ is $\sim$2700\text{\AA}. This allows for a better comparison with the UV ($u'$) surface brightness distribution for supernova environments, as reported by \citet{2012ApJ...759..107K} than made by \citet{2024MNRAS.527L..47C} with $F555W$. The Galactic extinction-corrected $F336W$ surface brightness is in the faintest $\sim$10\% for local supernova values; this is therefore faint but not unprecedented. 

\begin{table}
\centering
	\caption{{\em HST} magnitudes $m$, and their uncertainties $\delta m$, for the second epoch of AT\,2023fhn imaging at $\sim200$ days (Table \ref{tab:hstdata}). Upper limits are given at 3$\sigma$. In all six filters, two photometry methods are listed - aperture photometry with median background estimation, and aperture photometry with annulus background estimation. Two aperture sizes (and hence enclosed energy corrections) are given in each case.}
	\label{tab:phot_table}
	\begin{tabular}{lccccr} 
		\hline
		\hline
		Filter & Method & Bkg. & Aper. & m & $\delta$m \\
		\hline
            F225W	 & 	{\sc photutils}	 & 	Median	 & 	0.2$''$	 & 	>26.1	 & 	 -	\\
            F225W	 & 	{\sc photutils}	 & 	Annulus	 & 	0.2$''$	 & 	>26.1	 & 	 -	\\
            F225W	 & 	{\sc photutils}	 & 	Median	 & 	0.4$''$	 & 	>25.4	 & 	 -	\\
            F225W	 & 	{\sc photutils}	 & 	Annulus	 & 	0.4$''$	 & 	>25.5	 & 	 -	\\
		\hline
            F336W	 & 	{\sc photutils}	 & 	Median	 & 	0.2$''$	 & 	>26.6	 & 	 -	\\
            F336W	 & 	{\sc photutils}	 & 	Annulus	 & 	0.2$''$	 & 	>26.6	 & 	 -	\\
            F336W	 & 	{\sc photutils}	 & 	Median	 & 	0.4$''$	 & 	>25.9	 & 	 -	\\
            F336W	 & 	{\sc photutils}	 & 	Annulus	 & 	0.4$''$	 & 	>25.9	 & 	 -	\\
		\hline
            F555W	 & 	{\sc photutils}	 & 	Median	 & 	0.2$''$	 & 	26.9	 & 	0.2	\\
            F555W	 & 	{\sc photutils}	 & 	Annulus	 & 	0.2$''$	 & 	27.1	 & 	0.3	\\
            F555W	 & 	{\sc photutils}	 & 	Median	 & 	0.4$''$	 & 	25.8	 & 	0.2	\\
            F555W	 & 	{\sc photutils}	 & 	Annulus	 & 	0.4$''$	 & 	25.6	 & 	0.1	\\
		\hline
            F763M	 & 	{\sc photutils}	 & 	Median	 & 	0.2$''$	 & 	>26.0	 & 	 -	\\
            F763M	 & 	{\sc photutils}	 & 	Annulus	 & 	0.2$''$	 & 	>26.0	 & 	 -	\\
            F763M	 & 	{\sc photutils}	 & 	Median	 & 	0.4$''$	 & 	>25.3	 & 	 -	\\
            F763M	 & 	{\sc photutils}	 & 	Annulus	 & 	0.4$''$	 & 	25.0	 & 	0.3	\\
		\hline
            F814W	 & 	{\sc photutils}	 & 	Median	 & 	0.2$''$	 & 	26.4	 & 	0.2	\\
            F814W	 & 	{\sc photutils}	 & 	Annulus	 & 	0.2$''$	 & 	26.5	 & 	0.3	\\
            F814W	 & 	{\sc photutils}	 & 	Median	 & 	0.4$''$	 & 	25.3	 & 	0.2	\\
            F814W	 & 	{\sc photutils}	 & 	Annulus	 & 	0.4$''$	 & 	25.2	 & 	0.2	\\
		\hline
            F845M	 & 	{\sc photutils}	 & 	Median	 & 	0.2$''$	 & 	>25.6	 & 	 -	\\
            F845M	 & 	{\sc photutils}	 & 	Annulus	 & 	0.2$''$	 & 	>25.6	 & 	 -	\\
            F845M	 & 	{\sc photutils}	 & 	Median	 & 	0.4$''$	 & 	>24.9	 & 	 -	\\
            F845M	 & 	{\sc photutils}	 & 	Annulus	 & 	0.4$''$	 & 	>24.9	 & 	 -	\\
		\hline
	\end{tabular}
\end{table}

\subsection{Global host properties}\label{sec:hostsed}
We next consider how the overall host galaxy properties compare with the local environment of AT\,2023fhn, and how they compare with the hosts of other LFBOTs. To do this, we perform spectral energy distribution (SED) fitting of the integrated light of the host. By host, we refer to the spiral and satellite galaxy together, since their proximity likely results in interactions (e.g. tidal) and therefore the two galaxies can be considered as one interacting system. Furthermore, the two galaxies are not spatially resolved in ground-based imaging (e.g. PanSTARRS), which we used to add photometric points to the SED. 

\begin{figure*}
   \centering
   \includegraphics[width=\textwidth]{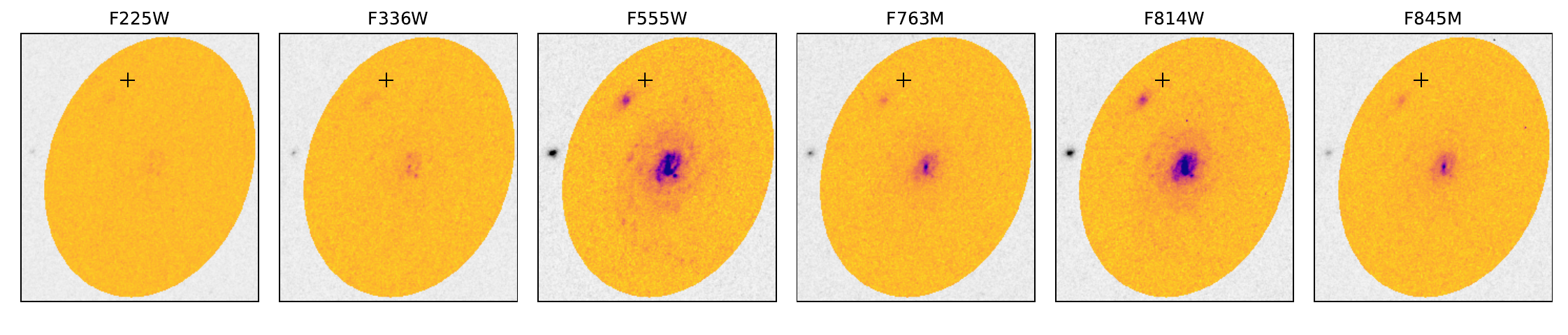}
      \caption{HST imaging of the galaxy hosting AT\,2023fhn in the six epoch 2 filters. Pixels within 1.5\,R$_{\rm petro}$ of the spiral galaxy centroid are selected as associated with the host, and given an orange-purple colourmap (see text for details). This region fully encompasses the satellite galaxy. The location of AT\,2023fhn is marked with a $+$ sign. The image cutouts are 13\,arcsec on each side. North is up and east is left.}
         \label{fig:hostSEDs}
   \end{figure*}

We attempt to collect as close to 100\% of the galaxy light from HST photometry as possible. We measure the Petrosian radius $R_{\rm petro}$ \citep{1976ApJ...209L...1P} of the spiral galaxy with the {\sc statmorph} package \citep[][$\eta=0.2$]{2019MNRAS.483.4140R} and adopt 1.5$R_{\rm petro}$ as a radius that encloses $\sim$100\% of the flux \citep[e.g.][]{2003ApJS..147....1C}. We account for the projected ellipticity and orientation of the galaxy using the {\sc ellip} and {\sc theta} outputs. A pixel mask is produced using these parameters as measured from the $F555W$ image, and applied to the other HST images, as shown in Figure \ref{fig:hostSEDs}. The flux within the mask is summed, and background subtraction (as for the local environment measurements above) uses the sigma-clipped median background, scaled for the number of pixels in the mask. Repeating the procedure for the satellite galaxy produces a 1.5$R_{\rm petro}$ pixel mask that lies entirely within the spiral's mask. We therefore use spiral pixel mask alone as it captures $\sim$100\% of the flux from both galaxies. Uncertainties are determined using the standard deviation of background pixels outside the galaxy mask, again using background estimator, the total background noise on the measurement is then $\sqrt{N} \times \sigma_{bg}$. To investigate the effect of background variations, we split each image into regions of $400 \times 400$ pixels - the approximate size of the galaxy mask - and determine the median background level in each region. The standard deviation of these, i.e. the uncertainty in the background level, is added in quadrature as an additional source of error on our HST galaxy photometry.

To supplement the HST data we add host photometry from archival catalogues. For additional optical points we use PanSTARRS data release 2 \citep{2016arXiv161205560C}. We use the catalogued Kron magnitudes \citep[][in $g$, $r$, $i$, $z$ and $y$]{1980ApJS...43..305K}, which capture $\sim$90\% of the light of extended sources, and increase the fluxes by a further 10\% to approximate the $\sim$100\% flux value \footnote{\url{https://outerspace.stsci.edu/display/PANSTARRS/PS1+Kron+photometry+of+extended+sources}}. The Kron radii for the spiral (4.41, 4.62, 4.36, 3.33 and 2.89\,arcsec in $g,r,i,z,y$ respectively) extend past the position of the satellite in $g,r,i$, so the system can be considered blended in these filters. In $y$ and $z$ this is not the case, so they likely underestimate the flux from the combined galaxy-satellite system, and indeed these points lie a few sigma below the best-fit spectrum when they are included in the fit. We therefore remove the $y$ and $z$ filters from the fitting. We also add far-UV and near-UV photometry from GALEX \citep{2003SPIE.4854..336M}, plus W1, W2 and W3 detections from WISE. Effective wavelengths for these filters are taken from \citet{2012ApJ...750...99T}. The spiral and satellite cannot be separated at the spatial resolution of these surveys, and neither galaxy is detected in 2MASS. The full list of photometry used to performed SED fitting is provided in Table \ref{tab:hostphot}.

To perform SED fitting we use {\sc prospector} \citep{2017ApJ...837..170L,2021ApJS..254...22J}, which makes use of FSPS \cite[Flexible Stellar Population Synthesis][]{2009ApJ...699..486C,2010ApJ...712..833C} and Python-FSPS \citep{2023zndo..10026684J}. For the Markov Chain Monte Carlo (MCMC) implementation we use {\sc emcee} \citep{2013PASP..125..306F}. We again use BPASS \citep[Binary Population and Spectral Synthesis v2.1,][]{2017PASA...34...58E,2018MNRAS.479...75S} for the spectral models. Before being passed to {\sc prospector}, the input photometry is corrected for Galactic extinction (as described in Section \ref{sec:local}). We fit four parameters: the stellar mass $M_{\star}$, intrinsic (local to the transient) extinction $A_{\rm V}$, population age $t_{\rm age}$ and the timescale for an exponentially declining star-formation history $\tau$. The redshift is fixed at $z=0.238$, and the luminosity distance at $D_{\rm L} = 1192$\,Mpc. 

\begin{figure*}
\centering
\includegraphics[width=0.95\textwidth]{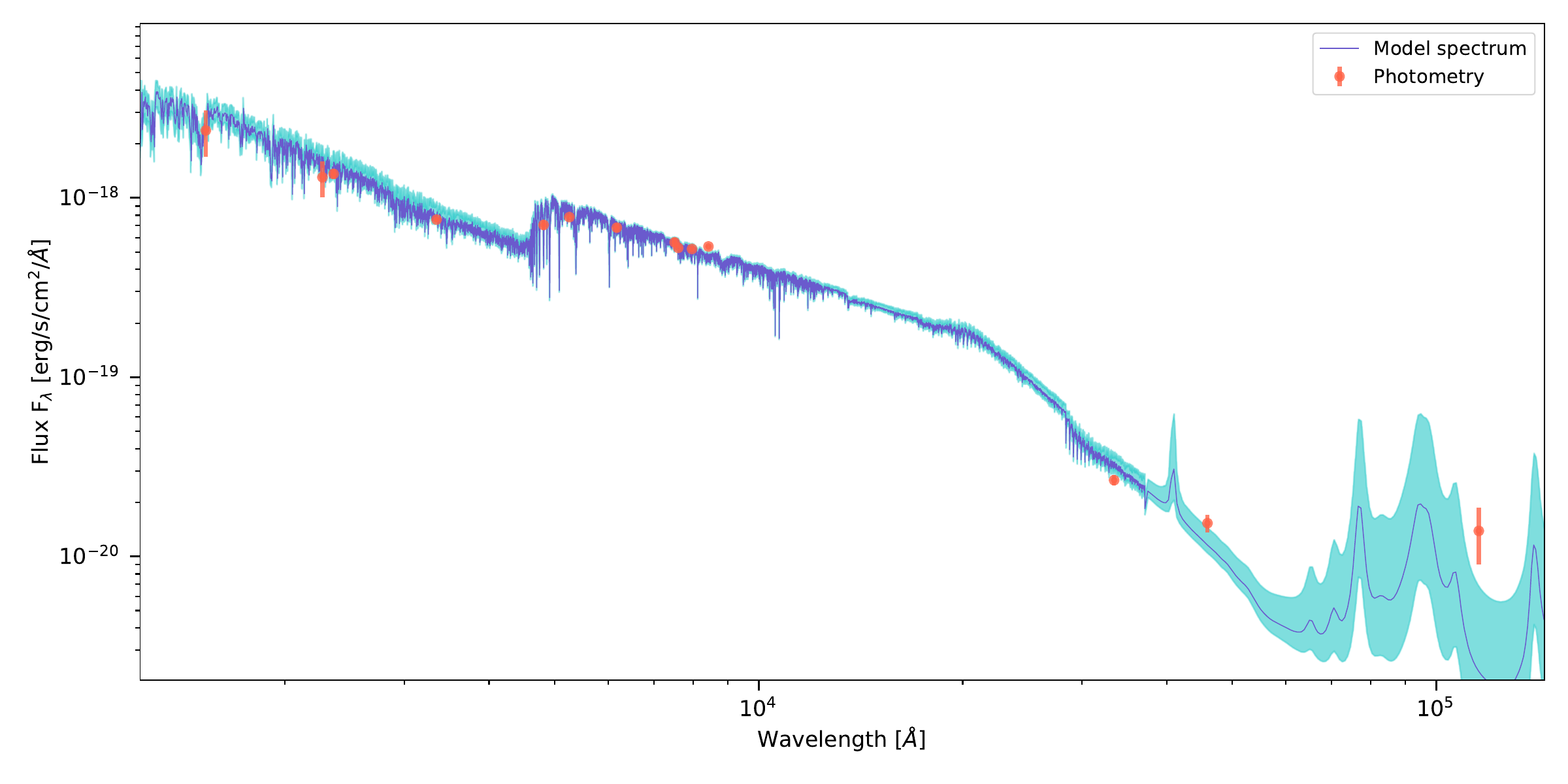}
\caption{Host galaxy photometry and best-fit spectrum from {\sc prospector}. The model spectrum is red-shifted into the observer frame. A light blue shaded region encloses the 90\% confidence interval on the posterior flux distribution at each wavelength. The photometry is from GALEX, PanSTARRS, WISE and HST/WFC3 as listed in Table \ref{tab:hostphot}, and is corrected for Galactic extinction with the Python module {\sc extinction} at the filter effective wavelengths. The corresponding galaxy properties are listed in Table \ref{tab:hostresults}. }
\label{fig:SED}
\end{figure*}

We run the MCMC with 128 walkers and 512 iterations; the full list of MCMC set-up parameters and joint posterior distributions (in the form of a corner plot) are provided in Appendix \ref{app:A}. The maximum a posterior (MAP) spectrum is shown in Figure \ref{fig:SED}, with the associated properties from the posterior distribution listed in Table \ref{tab:hostresults}. Thus far, the metallicity $Z$ has been fixed at half-Solar, based on the approximate mass of $10^{10}$\,M$_{\odot}$ and the mass-metallicity relation \citep{2004ApJ...613..898T,2005MNRAS.362...41G}. A similar table containing the results when metallicity is allowed to vary is also provided in Appendix \ref{app:A}. In this case, the mass and SFR are similar, such that fixing $Z$ at a more realistic value does not change our results in a qualitative sense. In the delayed-$\tau$ model, the current star-formation rate (SFR) is proportional to $(t/ \tau)e^{(-t/ \tau)}$. The absolute value is obtained by normalisation with respect to the mass formed, yielding a SFR of 5.5\,M$_{\odot}$\,yr$^{-1}$. In Figure \ref{fig:SED}, the F845M point lies above the best fit spectrum, possibly due to the presence of H$\alpha$ emission in this medium-width filter. Using the offset between the F845M point and the best-fit spectrum, and assuming the offset is dominated by H$\alpha$ emission, we estimate the line flux as $\sim$1$\times10^{-19}$erg\,s$^{-1}$\,cm$^{-2}$\,$\AA^{-1}$, which corresponds to a SFR of $\sim$1\,M$_{\odot}$yr$^{-1}$ \citep{1994ApJ...435...22K}. However, since the F845M point is only offset by $\sim$3$\sigma$, and given the uncertainty in the continuum level, we use continue to use the {\sc prospector} derived SFR as the fiducial value.

\begin{figure}
   \centering
   \includegraphics[width=\columnwidth]{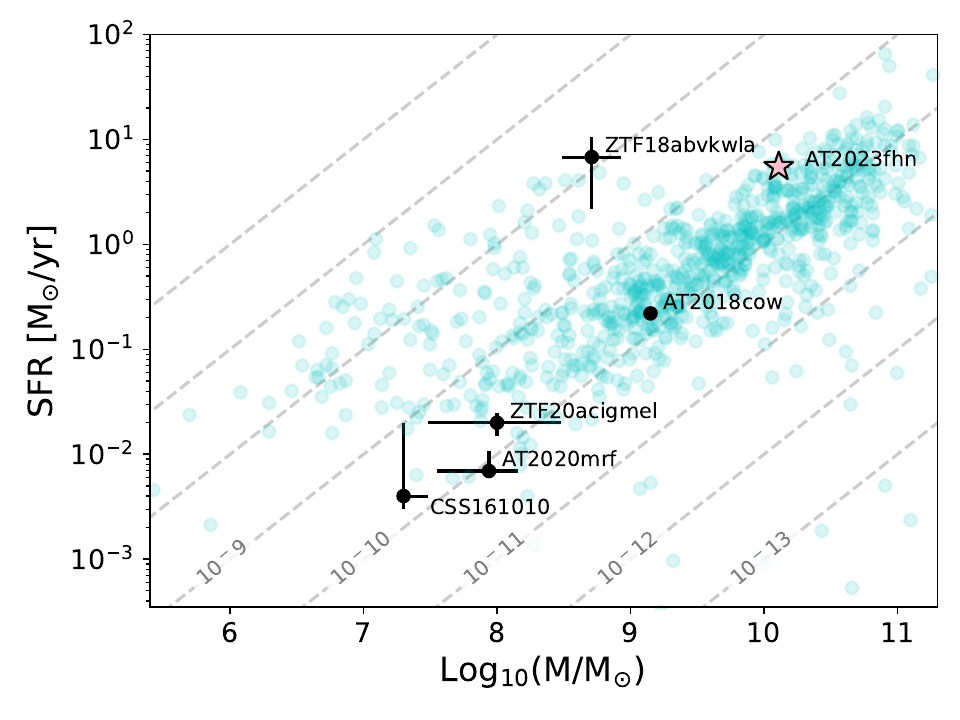}
      \caption{Stellar mass versus SFR for LFBOT host galaxies, including AT\,2023fhn. Other LFBOT host data are from \citet[][AT\,2018cow]{2019MNRAS.484.1031P}, \citet[][ZTF\,18abvkwla]{2020ApJ...895...49H}, \citet[][CSS161010]{2020ApJ...895L..23C} and \citet[][ZTF\,20acigmel]{2021MNRAS.508.5138P}. Lines of constant specific star formation rate (sSFR/yr$^{-1}$) are drawn in grey. The core-collapse supernova host galaxy sample of \citet{2021ApJS..255...29S} is plotted as transparent cyan points.}
         \label{fig:massSFR}
   \end{figure}

In summary, the galaxy pair is dominated by a fairly typical star-forming spiral, but is perhaps notable for the likely presence of tidal interactions between the spiral and its satellite. In Figure \ref{fig:massSFR} we plot its mass versus SFR, comparing with the host galaxies of previous LFBOTs. The galaxy has a high SFR and mass for LFBOT hosts, lying slightly above average in terms of specific star formation rate (sSFR), but well below the sSFR of the host of ZTF\,18abvkwla.

\begin{table}
	\centering
	\caption{Host galaxy photometry used for SED fitting. All magnitudes are in the AB system, and before Galactic extinction correction. The filter effective wavelengths and Galactic extinction at that wavelength - assuming E(B-V)=0.0254, R$_{\rm V}=3.1$ and a \citet{1999PASP..111...63F} extinction law - are also listed. We increase the PanSTARRS fluxes by 10\% over the values below, as described in the text.}
	\label{tab:hostphot}
	\begin{tabular}{cccccc} 
		\hline
		\hline
		  Filter & Source & m & err & $\lambda_{\rm eff}$ [\text{\AA}] & A($\lambda$) \\
		\hline
            FUV & GALEX & 20.93 & 0.31 & 1548.85 & 0.20\\
            NUV & GALEX & 20.74 & 0.25 & 2303.37 & 0.22 \\
            F225W & HST & 20.60 & 0.07 & 2358.70 & 0.20 \\
            F336W & HST & 20.40 & 0.03 & 3359.11 & 0.13 \\
            $g$ & PS & 19.70 & 0.01 & 4810.00 & 0.09 \\
            F555W & HST & 19.34 & 0.01 & 5235.33 & 0.08\\
            $r$ & PS & 19.17 & 0.01 & 6170.00 & 0.07 \\
            $i$ & PS & 18.93 & 0.01 & 7520.00 & 0.05 \\
            F763M & HST & 18.93 & 0.01 & 7602.85 & 0.05 \\
            F814W & HST & 18.84 & 0.01 & 7954.84 & 0.04 \\
            F845M & HST & 18.68 & 0.02 & 8430.20 & 0.04 \\
            W1 & WISE & 18.91 & 0.07 & 33526.00 & 0.00 \\
            W2 & WISE & 18.82 & 0.13 & 46028.00 & 0.00 \\
            W3 & WISE & 16.92 & 0.38 & 115608.00 & 0.00 \\
		\hline
    \end{tabular}
\end{table}

\begin{table}
	\centering
	\caption{Host galaxy properties derived from {\sc prospector} SED fitting. The median values from the marginalised posterior distributions are quoted, with uncertainties bounding the 68\% confidence interval on each parameter.}
	\label{tab:hostresults}
	\begin{tabular}{cc} 
		\hline
		\hline
		  Host property & Value \\
		\hline
            $M_{\star}$ / M$_\odot$ & $(1.29^{+0.08}_{-0.07}) \times 10^{10}$ \\
            SFR / M$_\odot$\,yr$^{-1}$  & 5.5$^{+1.1}_{-0.7}$ \\
            $A_{{\rm V}}$  & 0.20$\pm0.07$ \\
            $t_{{\rm age}}$/Gyr  & 1.68$^{+0.29}_{-0.20}$ \\
            $\tau$/Gyr & 0.66$^{+0.31}_{-0.13}$ \\
		\hline
    \end{tabular}
\end{table}

\section{Transient emission}\label{sec:transientemiss}~
\subsection{UV-optical}
\begin{figure}
   \centering
   \includegraphics[width=\columnwidth]{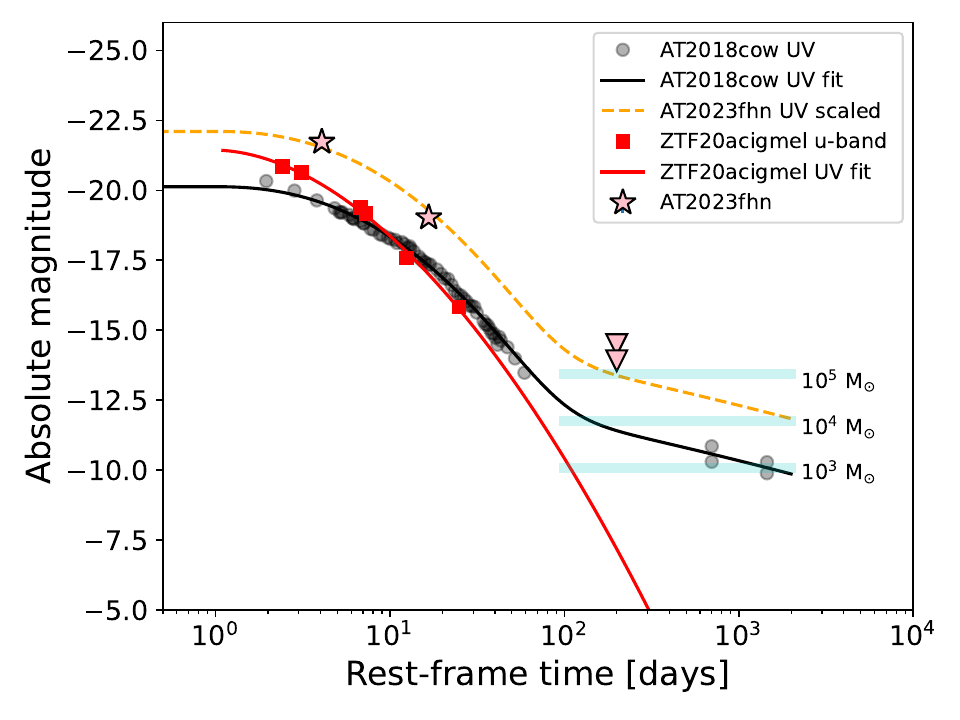}
      \caption{The UV data points for AT\,2023fhn, compared with the UV light-curves of AT\,2018cow \citep[Early $u$-band and late-time F222W and F336W,][]{2018ApJ...865L...3P,2019MNRAS.484.1031P,2023MNRAS.525.4042I} and ZTF\,20acigmel \citep{2021MNRAS.508.5138P}. A light-curve fit to the AT\,2018cow data is increased in luminosity to intercept the sole early-time AT\,2023fhn UV point, the subsequent F336W limit at $\sim$112 rest-frame days lies just above the expected UV magnitude at this epoch, assuming identical evolution to the Cow. A similar fit is made for the early-time ZTF\,20acigmel points. Cyan horizontal bands show the expected UV absolute magnitudes at late times for accretion discs around intermediate mass black holes of different masses, following a tidal disruption event \citep{2024MNRAS.527.2452M}.}
         \label{fig:optuv}
   \end{figure}

We now compare the UV-optical constraints on AT\,2023fhn's light-curve with previous LFBOTs. All times used in Section \ref{sec:transientemiss} are in the rest-frames of the LFBOTs. Comparison data are corrected for Galactic extinction of $E(B-V)$=0.08 \citep[AT\,2018cow,][]{2018ApJ...865L...3P} and $E(B-V)$=0.07 \citep[ZTF\,20acigmel,][]{2021MNRAS.508.5138P}, their UV light-curves (in absolute magnitude) are compared with AT\,2023fhn in Figure \ref{fig:optuv}. We fit the light-curve of AT\,2018cow in 2 phases, early ($<200$~d) and late-time, with a fit of the form $M=a \log(t)^{b}+c$. For the fit to AT\,2018cow, we assume that the late-time UV is dominated by residual transient emission \citep[][]{2022MNRAS.512L..66S,2023MNRAS.519.3785S,2023ApJ...955...43C,2023MNRAS.525.4042I}. We shift the AT\,2018cow best-fit up in absolute magnitude such that it lies between the early-time ATLAS $c$-band and FORS2 $u$-band AT\,2023fhn detections \citep{2023TNSAN..93....1H}. The extrapolated curve passes below the late-time HST F225W and F336W upper limits reported in this work. Another LFBOT with good UV photometric coverage is ZTF\,20acigmel, but here we consider only the early, pre-break phase due to a lack of late-time constraints. ZTF\,20acigmel starts brighter than AT\,2018cow and fades faster, whereas AT\,2023fhn is the most luminous LFBOT yet at UV-optical wavelengths. A final addition to Figure \ref{fig:optuv} are bands of constant UV absolute magnitude, corresponding to late-time emission from black holes of different masses in the tidal disruption event model of \citet{2024MNRAS.527.2452M}. This model yielded a black hole mass of $\sim$10$^{3}$\,M$_{\odot}$ for AT\,2018cow. Assuming that AT\,2023fhn had a similar light-curve shape to AT\,2018cow - which may not be the case for accretion events around black holes of different masses - the HST F336W point source upper limit for AT\,2023fhn tentatively constrains the accreting black hole mass in a TDE interpretation to $\lesssim 10^{5}$\,M$_{\odot}$. 

\subsection{X-ray}
\begin{figure}
   \centering
   \includegraphics[width=\columnwidth]{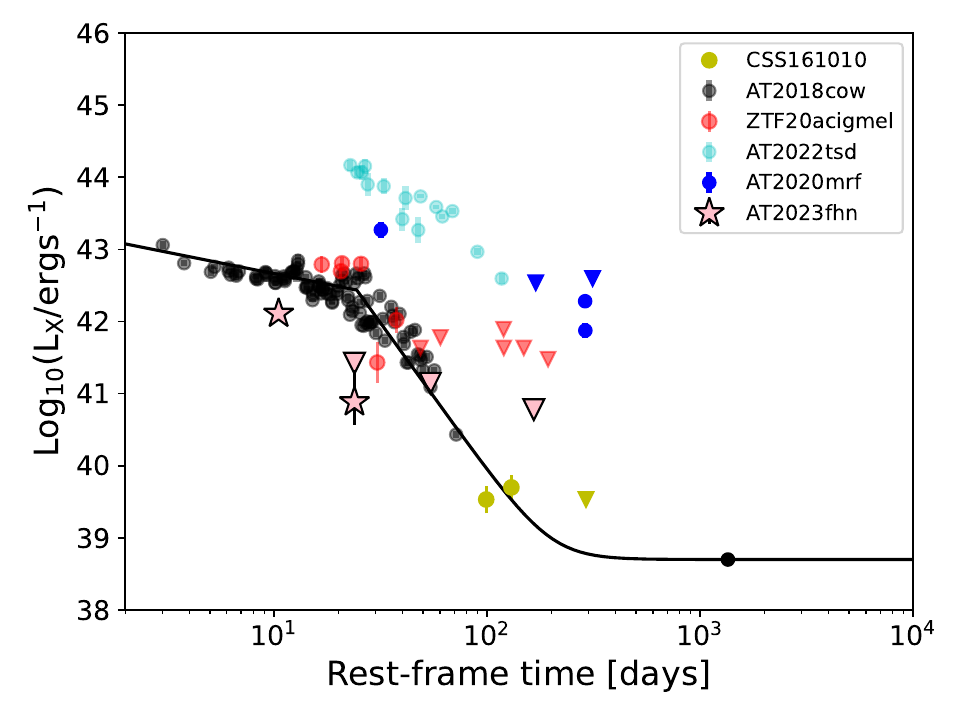}
      \caption{The X-ray light-curve of AT\,2023fhn compared with other LFBOTs. All data are in the $\sim$0.5-10\,keV range, are unabsorbed and from \citet[][AT\,2018cow]{2018MNRAS.480L.146R,2019MNRAS.487.2505K,2024ApJ...963L..24M}, \citet[][CSS161010]{2020ApJ...895L..23C}, \citet[][ZTF\,20acigmel]{2022ApJ...926..112B,2022ApJ...932..116H}, \citet[][AT\,2020mrf]{2022ApJ...934..104Y} and \citet[][AT\,2022tsd]{2023RNAAS...7..126M}. Given the marginal nature of the second AT\,2023fhn measurement, we also plot the 2$\sigma$ upper limit at this epoch. A broken-power law and late-time plateau interpretation of AT\,2018cow's light-curve is shown by the solid black line \citep{2024ApJ...963L..24M}.}
         \label{fig:xray}
   \end{figure}

Figure \ref{fig:xray} shows our X-ray observations of AT\,2023fhn, and the X-ray light-curves of other LFBOTs. The AT\,2018cow broken power-law and late-time plateau fit of \citet{2024ApJ...963L..24M} is also shown. AT\,2023fhn is the faintest LFBOT in X-rays at early times. Assuming a shallow decay initially, similar to AT\,2022tsd, ZTF\,20acigmel and AT\,2018cow, the break time can be - at the latest - similar to AT\,2018cow and ZTF\,20acigmel. Notwithstanding the small sample size, among the three LFBOTs with a clearly observed break in the X-ray light-curve (AT2018\,cow, ZTF20\,acigmel and AT\,2022tsd), brighter LFBOTs seem to transition to a steeper decay at later times. Assuming instead that epochs 1 and 2 are on the same phase of the light-curve, the decay index $n=2.1^{+0.7}_{-0.9}$ (where $L \propto t^{-n}$). Expectations for the X-ray decay rate are $t^{-1}$ (shock power), $t^{-2}$ (magnetar central engine) and $t^{-5/3}$ (fallback, i.e. a TDE). Overall, the detections and upper-limits are consistent with AT\,2023fhn behaving like a fainter version of previous LFBOTs in the X-ray band, and demonstrates that they can exhibit several orders of magnitude of variety in their X-ray luminosity.

\subsection{UV/X-ray ratio}
\begin{figure}
   \centering
   \includegraphics[width=\columnwidth]{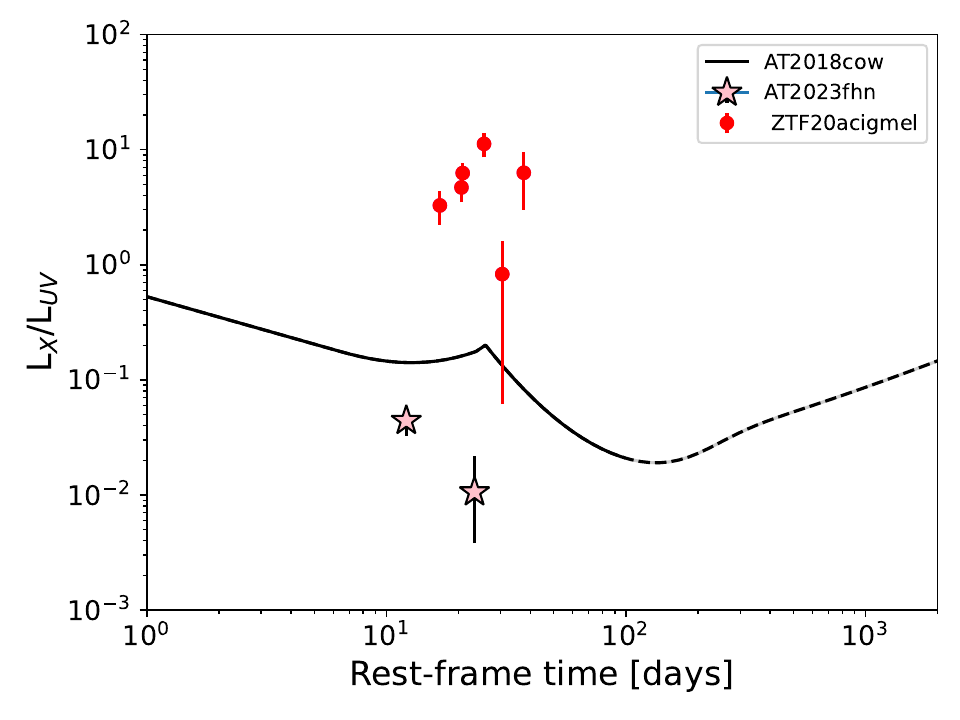}
      \caption{The ratio of X-ray to UV luminosity for LFBOTs AT\,2023fhn, AT\,2018cow and ZTF\,20acigmel. The AT\,2023fhn points use the dashed orange line in Figure \ref{fig:optuv} and the two {\em Chandra} X-ray detections (errorbars reflect the X-ray uncertainties only). The black curve is the ratio of the \citet{2024ApJ...963L..24M} X-ray light-curve fit (see the solid black line, Figure \ref{fig:xray}) and a broken power-law fit to the UV observations (the solid black line in Figure \ref{fig:optuv}). The small $k$-corrections are neglected in this comparison. The evolution past 200 days (drawn as a dashed line) is highly uncertain due to the sole X-ray detection. The ZTFacigmel data are from \citet{2022ApJ...932..116H,2022ApJ...926..112B}, where we have taken the ratio of X-ray points and the power-law fit to the ZTFacigmel UV-light-curve in Figure \ref{fig:optuv}. The uncertainties on these points again solely reflect the X-ray measurement uncertainties.}
         \label{fig:ratio}
   \end{figure}

Motivated by the fact that AT\,2023fhn appears to be the brightest LFBOT yet at UV-optical wavelengths, and the faintest in terms of X-ray luminosity, in Figure \ref{fig:ratio} we show the ratio of X-ray to UV luminosity for the 3 LFBOTs with such constraints. The data points for AT\,2023fhn take the X-ray detections at 12 and 23 rest-frame days, and the corresponding point on the shifted AT\,2018cow light-curve in Figure \ref{fig:optuv}. The uncertainties shown are exclusively from the X-ray observations. For AT\,2018cow, we take the ratio of the X-ray fit of \citet{2024ApJ...963L..24M} in Figure \ref{fig:xray}, and our fit to the UV light-curve fit in Figure \ref{fig:optuv}. Finally, for ZTF\,20acigmel we take the ratio of the X-ray luminosity with the UV light-curve fit at the same time. LFBOTs therefore exhibit at least $\sim3$ orders of magnitude in their X-ray/UV luminosity ratio, even at similar times in their evolution. This is plausibly a viewing angle effect. A qualitative prediction of tidal disruption models is a trade-off between UV-optical and X-ray luminosity as a function of viewing angle, where on-axis angles (which may also be aligned with a beamed outflow) would see a higher X-ray luminosity \citep{2018ApJ...859L..20D,2021ApJ...921...20H}. Differences in L$_{\rm X}$/L$_{\rm UV}$ are also expected for different black hole masses and spins, due to varying accretion disc formation rates \citep[which in turn affects the delay betweeen peak X-ray and UV/optical emission,][]{2020ApJ...889..166J}. However, a scenario in which the peak X-ray emission is delayed due to a delay in forming the inner accretion disc is hard to reconcile with the energetics and (variability) timescales of LFBOT emission, which demands energy input from a central engine and therefore active accretion \citep[e.g.][]{2019ApJ...871...73H,2019ApJ...872...18M}. Alternatively, the range of $L_{\rm x}$/$L_{\rm UV}$ could reflect differences in the circumstellar media, which we investigate in the following Section.

\subsection{Radio}\label{sec:radio_results}
Assuming that the radio emission is synchrotron-dominated with self-absorption - as we will see, the radio SED of AT\,2023fhn is consistent with this - and that the peak of the SED occurs at the synchtrotron self-absorption (SSA) frequency, we can estimate several shock parameters, and properties of the circumstellar medium. We follow the synchrotron self-absorption model of \citet{1998ApJ...499..810C} \citep[see also][]{2005ApJ...621..908S}. Adopting this framework for AT\,2023fhn is reasonable since this best fits other LFBOTs studied so far \citep[based on the brightness temperature, which precludes thermal emission, and the spectral shape, e.g.][]{2019ApJ...872...18M,2020ApJ...895L..23C,2020ApJ...895...49H,2021ApJ...912L...9N,2022ApJ...932..116H,2022ApJ...934..104Y,2022ApJ...926..112B}.

We fit the radio spectrum at $\sim$90 and $\sim$138 days ($\sim$70 and $\sim$110 rest-frame days) following \citet{1998ApJ...499..810C,2002ApJ...568..820G,2006ApJ...651..381C}. At a given time $t$ the radio SED has the form,
\begin{equation}\label{eq:eq1}
    F(\nu) = F_{\rm pk} \left[ \left( \frac{\nu}{\nu_{\rm pk}} \right)^{-s \beta_{1}} + \left( \frac{\nu}{\nu_{\rm pk}} \right)^{-s \beta_{2}} \right]^{-\frac{1}{s}} ,
\end{equation}
where $F(\nu)$ is the flux density, $F_{\rm pk}$ is the flux at the peak (break) frequency $\nu_{\rm pk}$ where the optically thick and thin power laws intersect, $s$ is a smoothing factor and $\beta_{1}$ and $\beta_{2}$ are spectral indices in the optically thick and thin regimes, respectively. In our case the cooling frequency $\nu_{\rm c}$ lies at higher frequencies than probed by our observations ($\sim$400-800\,GHz), where $\nu_{\rm c}$ is given by $18 \pi m_{e}ce)/(t^{2} \sigma_{t}^{2} B^{3} )$ \citep{2022ApJ...938...84D}. We therefore expect $F(\nu) \propto \nu^{-(p-1)/2}$ in the optically thin regime, where $p$ is the power law index of the electron energy distribution in the shock (i.e. the number $N$ of electrons with Lorentz factor $\gamma_{\rm e}$ goes as $N(\gamma_{\rm e}) \propto \gamma_{\rm e}^{-p}$).  

Using the {\sc scipy} {\sc curve\_fit} function, and working with rest-frame times and central frequencies throughout this section, we fit equation \ref{eq:eq1} to the $\sim$90\,day and $\sim$138\,day (observer frame) data. At 90 days we have 6 data points (5 detections, 1 upper limit, we combine the 87 and 95 day data for this epoch), and at 138 days we have 7 data points (6 detections, 1 upper limit). The data points (including limits, which use the rms as $\sigma$) are given a 1/$\sigma^{2}$ weighting (with {\sc sigma\_absolute = false}). There are 5 parameters to fit: $F_{\rm pk}$, $\nu_{\rm pk}$, $\beta_{1}$, $\beta_{2}$ and $s$. The best-fit values for these parameters and their uncertainties (calculated from the covariance matrix output by {\sc curve\_fit}) are listed in Table \ref{tab:results}. The optically-thin spectral index of -0.6 (138 days) yields an electron energy spectral index of $\sim$2.2, which is relatively shallow; 2.5 is expected from theory, while values closer to $\sim$3 are often measured in gamma-ray bursts, tidal disruption events and supernovae  \citep[e.g.][]{2006ApJ...651..381C,2024ApJ...971..185C}. Values from other LFBOTs are also in the range $\sim$2--3 \citep[][]{2019ApJ...872...18M,2020ApJ...895...49H,2020ApJ...895L..23C,2022ApJ...934..104Y,2022ApJ...926..112B}.

The peak flux F$_{\rm pk}$ and (rest-frame) frequency at the peak flux $\nu_{\rm pk}$ (at the intersection of the power-laws, rather than the fitted peak) allow us to estimate the radius of the shock, circumstellar density at that radius, and the CSM surface density parameter $A_{\star} \propto \dot{M} / v_{w}$ \citep[see][for a detailed description of the modelling assumptions]{2022ApJ...938...84D}. Following the formulism of \citet{1998ApJ...499..810C} (see also \citealt{2006ApJ...651..381C,2022ApJ...938...84D,2022ApJ...926..112B}), we first have the shock radius $R_{\rm p}$, given by,
\begin{equation}\label{eq:rp}
    R_{\rm p} = 4\times10^{14} \left( \frac{\epsilon_{\rm e}}{\epsilon_{\rm B}} \right)^{\frac{-1}{19}} \left( \frac{f}{0.5} \right)^{\frac{-1}{19}}
    \left( \frac{F_{\rm pk}}{{\rm (1+z)\,mJy}} \right)^{\frac{9}{19}} \left( \frac{D_{\theta}}{{\rm Mpc}} \right)^{\frac{18}{19}} \left( \frac{\nu_{\rm pk}}{5~{\rm GHz}} \right)^{-1} {\rm cm},
\end{equation}
where D$_{\theta}$ is the angular diameter distance, and $\epsilon_{e}$ and $\epsilon_{B}$ are the fraction of the shock energy in electrons and in the magnetic field, respectively. The average shock velocity can then be calculated as $R_{\rm p}/t_{\rm obs} = \Gamma \beta c$, where $\beta=v/c$, $\Gamma$ is the Lorentz factor and $t_{\rm obs}$ is the rest-frame observation time. Next we have, for the internal magnetic field $B$,
\begin{equation}
    B = 1.1 \left( \frac{\epsilon_{\rm e}}{\epsilon_{\rm B}} \right)^{\frac{-4}{19}} \left( \frac{f}{0.5} \right)^{\frac{-4}{19}}
    \left( \frac{F_{\rm pk}}{{\rm (1+z)\,mJy}} \right)^{\frac{-2}{19}} \left( \frac{D_{\theta}}{{\rm Mpc}} \right)^{\frac{-4}{19}} \left( \frac{\nu_{\rm pk}}{5~{\rm GHz}} \right) {\rm G}
\end{equation}
and for the wind density (the mass loss rate $\dot{M}$ over the wind velocity),
\begin{multline}\label{eq:mdotvw}
    \frac{\dot{M}}{v_{\rm w}} \left( \frac{1000~{\rm kms^{-1}}}{10^{-4} ~M_{\odot}yr^{-1}} \right)  = 2.5\times10^{-5} \left( \frac{1}{\epsilon_{\rm B}} \right) \left( \frac{\epsilon_{\rm e}}{\epsilon_{\rm B}} \right)^{\frac{-8}{19}} \left( \frac{f}{0.5} \right)^{\frac{-8}{19}} \times \\
    \left( \frac{F_{\rm pk}}{{\rm (1+z)\,Jy}} \right)^{\frac{-4}{19}} \left( \frac{D_{\theta}}{{\rm Mpc}} \right)^{\frac{-8}{19}} \left( \frac{\nu_{\rm pk}}{5~{\rm GHz}} \right)^{2} \left( \frac{t_{\rm pk}}{{\rm days}} \right)^{2}
\end{multline}
Under the assumption that the CSM is dominated by fully ionised hydrogen, the electron number density can be related to $\dot{M}/v_{w}$ by $n_{e} = \dot{M}/(4 \pi m_{p} r^{2} v_{w})$ - where $m_{p}$ is the proton mass - so that,
\begin{multline}\label{eq:ne}
    n_{\rm e} = 1.02418 \left( \frac{1}{\epsilon_{\rm B}} \right) \left( \frac{\epsilon_{\rm e}}{\epsilon_{\rm B}} \right)^{\frac{-6}{19}} \left( \frac{f}{0.5} \right)^{\frac{-6}{19}} \times \\
    \left( \frac{F_{\rm pk}}{{\rm (1+z)\,Jy}} \right)^{\frac{-22}{19}} \left( \frac{D_{\theta}}{{\rm Mpc}} \right)^{\frac{-44}{19}} \left( \frac{\nu_{\rm pk}}{5~{\rm GHz}} \right)^{4} \left( \frac{t_{\rm pk}}{{\rm days}} \right)^{2} {\rm cm}^{-3}
\end{multline}
Additionally we have, for the internal shock energy $U = U_{\rm B} / \epsilon_{\rm B}$,
\begin{multline}\label{eq:U}
    U = 1.859\times10^{46} \left( \frac{1}{\epsilon_{\rm B}} \right) \left( \frac{\epsilon_{\rm e}}{\epsilon_{\rm B}} \right)^{\frac{-11}{19}} \left( \frac{f}{0.5} \right)^{\frac{8}{19}} \times \\
    \left( \frac{F_{\rm pk}}{{\rm (1+z)\,Jy}} \right)^{\frac{23}{19}} \left( \frac{D_{\theta}}{{\rm Mpc}} \right)^{\frac{46}{19}} \left( \frac{\nu_{\rm pk}}{5~{\rm GHz}} \right)^{-1} {\rm erg}
\end{multline}
We assume equipartition ($\epsilon_{e}$=$\epsilon_{B}$=1/3), where the magnetic energy density, the energy density in electrons and the energy density in protons contribute equally as destinations for the converted kinetic energy in the shock. We further assume $f=0.5$ for the filling factor. If the emission region is modelled as a disc of radius $R$ and thickness $S$ on the sky, whose volume is $\pi R^{2} S$, an equivalent spherical volume can be given by $4/3 \pi R^{3}$. The filling factor is the fraction of this equivalent spherical volume producing emission \citep{1998ApJ...499..810C}.

\begin{figure*}
   \centering
   \includegraphics[width=0.49\textwidth]{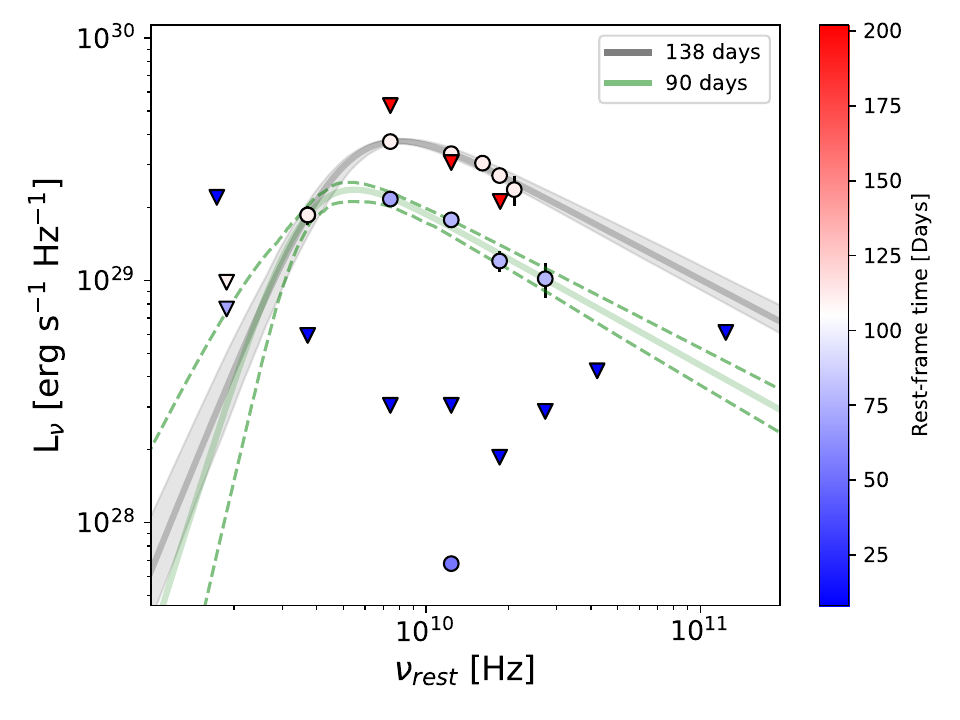}
   \includegraphics[width=0.49\textwidth]{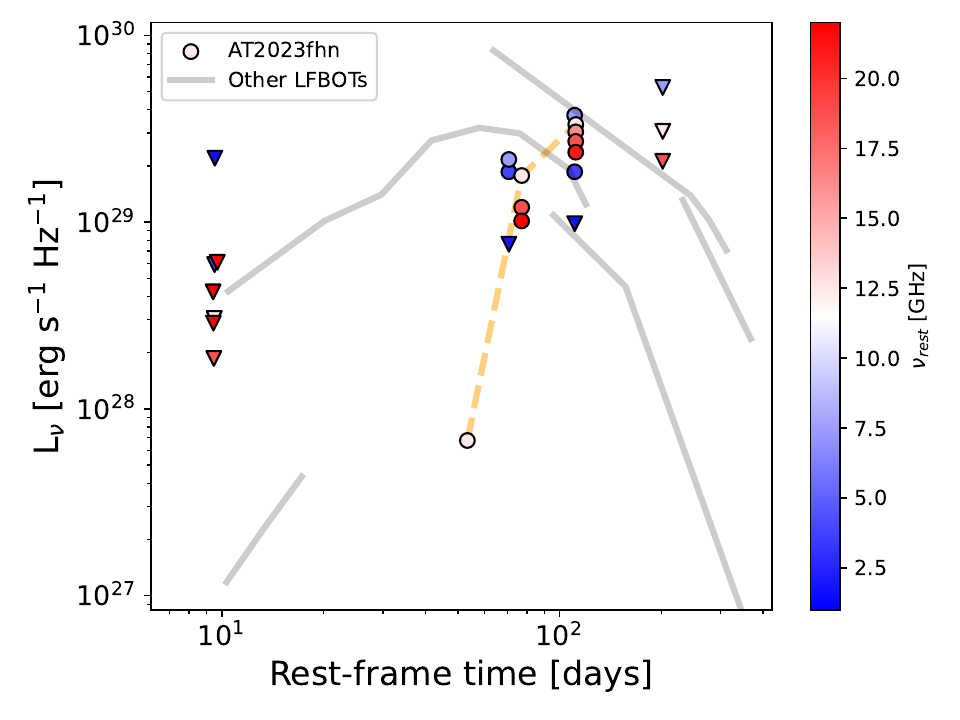}
      \caption{Radio observations of AT\,2023fhn (see Table \ref{tab:radio_table}) placed in the context of other LFBOTs. Upper limits from NOEMA and the VLA as reported by \citet{2023TNSAN.100....1H} and \citet{2023TNSAN.174....1H} are also shown. Left: radio SED for AT\,2023fhn, with a broken-power law fit to the $t=138$ day (grey line) and $t=90$ day (green line) data. Data point rest-frame times are indicated by the colourbar. The 90\% confidence regions on the fits are shown by light grey shading (138 days) and dashed green lines (90 days). Note that there are overlapping data points at $\sim$3\,Ghz (i.e. a 90 day point with similar luminosity is obscured by the 138 day point). Right: radio light-curve for AT\,2023fhn (central frequencies indicated by the colourbar) and other LFBOTs. The 10\,GHz AT\,2023fhn detections are connected by a dashed orange line, to aid the eye in comparing with other LFBOTs. Data for the other LFBOTs, all at (10$\pm$2)\,GHz, are from \citet[][AT\,2018cow]{2019ApJ...871...73H,2019ApJ...872...18M}, \citet[][ZTF\,18abvkwla]{2020ApJ...895...49H}, \citet[][CSS161010]{2020ApJ...895L..23C}, \citet{2022ApJ...926..112B}, \citet[][ZTF\,20\,acigmel]{2022ApJ...932..116H} and \citet[][AT\,2020mrf]{2022ApJ...934..104Y}.}
         \label{fig:radio}
   \end{figure*}

\begin{figure*}
   \centering
   \includegraphics[width=0.49\textwidth]{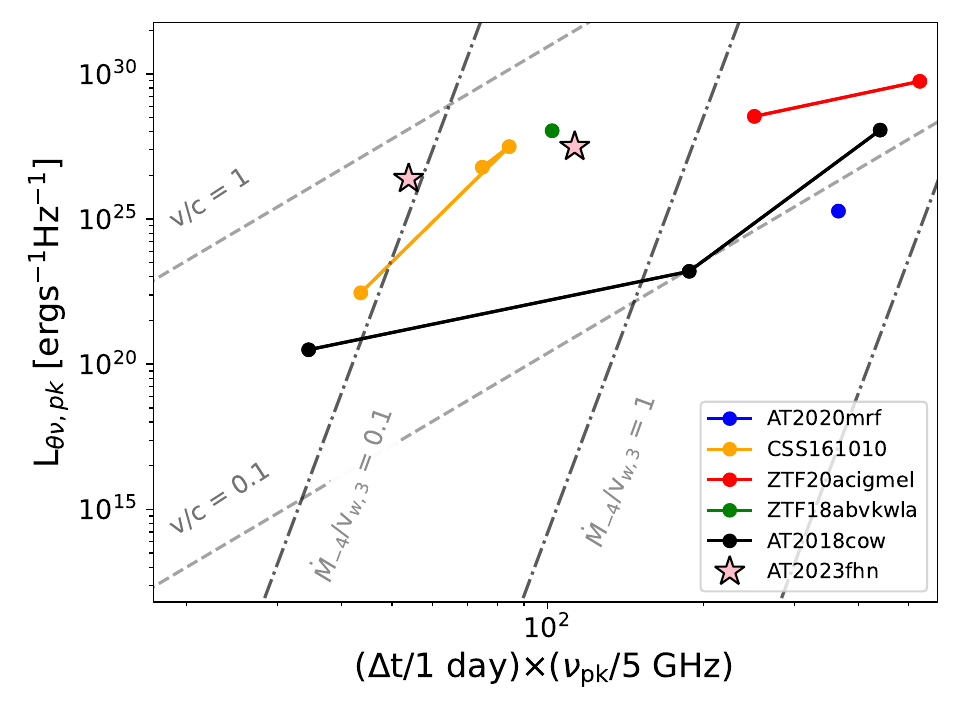}
   \includegraphics[width=0.49\textwidth]{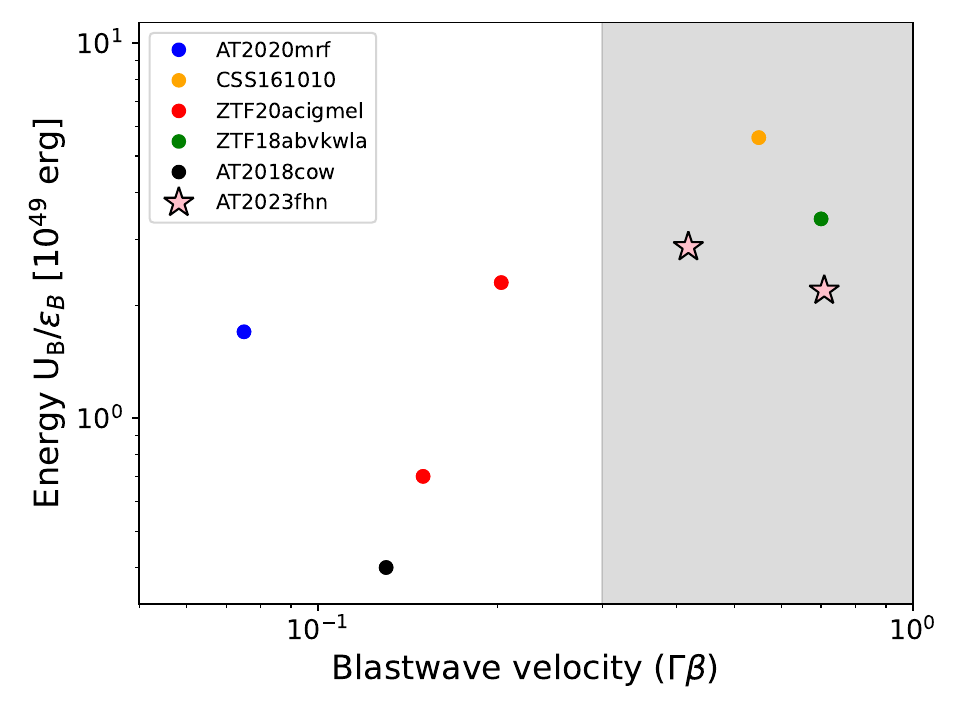}
      \caption{Left: the product of rest-time $\Delta t$ and the rest-frame peak frequency ${\nu}_{\rm pk}$ at that time, versus the peak radio spectral luminosity $L_{\rm \theta \nu,pk} = L_{\rm \nu,pk} / (1+z)^{3}$, with lines of constant $\dot{M}/V_{w}$ (in units of 10$^{-4}$\,M$_{\odot}$\,yr$^{-1}$ / 1000\,km\,s$^{-1}$) and lower limits on the blast-wave velocity shown. Both AT\,2023fhn epochs are shown, we note that constraints on the later point (138 days) are stronger and the parameters at this epoch better constrained. All comparison data points adopt $\epsilon_{e}=\epsilon_{B}=1/3$. Right: lower limits on the average blast-wave velocity (blast-wave radius over the rest-frame time) in units of $c$ versus the internal energy of the shock $U = U_{B}/\epsilon_{B}$. Other LFBOT data are taken from \citet[][AT\,2018cow]{2019ApJ...871...73H,2019ApJ...872...18M,2021ApJ...912L...9N}, \citet[][ZTF\,18abvkwla]{2020ApJ...895...49H}, \citet[][CSS161010]{2020ApJ...895L..23C}, \citet{2022ApJ...926..112B};\citet[][ZTF\,20\,acigmel]{2022ApJ...932..116H} and \citet[][AT\,2020mrf]{2022ApJ...934..104Y}. The mildly-relativistic regime is shaded. We have scaled the \citet{2022ApJ...926..112B} ZTF\,20\,acigmel point following equations \ref{eq:rp} and \ref{eq:U} to align with our assumption of equipartition.}
         \label{fig:synchfits}
   \end{figure*}

\begin{figure}
   \centering
   \includegraphics[width=\columnwidth]{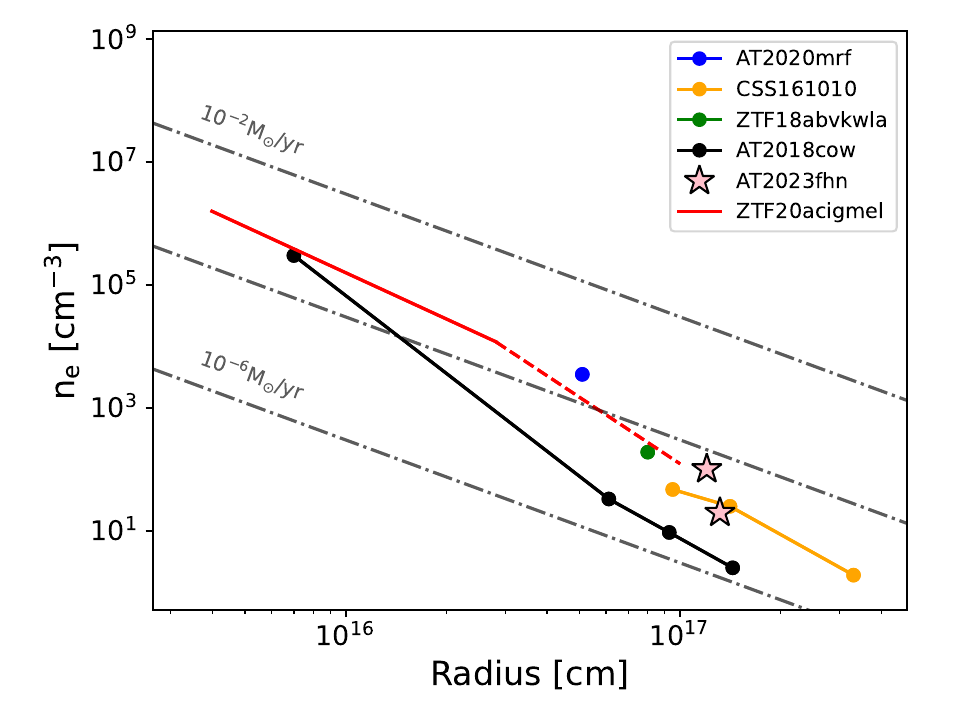}
      \caption{Circumstellar density n$_{\rm e}$ at the radius of the shock R$_{\rm p}$ for AT\,2023fhn and previous LFBOTs. The densities for ZTF\,20\,acigmel \citep[][Table 6]{2022ApJ...926..112B} have been reduced by a factor $\sim$16 to align them with the equipartition $\epsilon_{e}=\epsilon_{B}=1/3$ assumption used for all other measurements (for reference, see how equation \ref{eq:ne} scales with $\epsilon_{e}$ and $\epsilon_{B}$). As in Figure \ref{fig:synchfits}, both AT\,2023fhn epochs are shown, where the best constraints are from the epoch at 138 days. Other LBFOT results are from \citet[][AT\,2018cow]{2019ApJ...871...73H,2021ApJ...912L...9N}, \citet[][ZTF\,18abvkwla]{2020ApJ...895...49H}, \citet[][CSS161010]{2020ApJ...895L..23C}, \citet[][ZTF\,20\,acigmel, their fit to n$_{e}$(r) is adopted, shown as a red solid/dash line either side of a possible break]{2022ApJ...926..112B} and \citet[][AT\,2020mrf]{2022ApJ...934..104Y}. Lines of constant $\dot{M}$ are shown for $v_{w}=1000$\,km\,s$^{-1}$. Note that due to different assumptions in the modelling, the densities and/or mass loss rates derived between different authors and objects can differ by up to a factor 5 \citep{2022ApJ...938...84D}.}
         \label{fig:CSM}
   \end{figure}

We list the inferred properties of AT\,2023fhn's blast-wave in Table \ref{tab:results}. Results for the fiducial parameters of $\epsilon_{e}=0.1$ and $\epsilon_{B}=0.01$ are also listed. These properties are compared with other LFBOTs in Figures \ref{fig:radio}, \ref{fig:synchfits} and \ref{fig:CSM}. In LFBOTs the expanding blast-wave typically shows a SSA spectrum that decreases in peak flux and frequency over time. However we note that AT2023fhn shows an increase in peak flux between $\sim90$ and $\sim138$ days post explosion. A similar increase was seen in CSS161010 between 69 and 99 days post explosion \citep{2020ApJ...895L..23C}. This could potentially be caused by an increase in density, or inhomogeneities in the CSM, but we are not able to test this scenario given our weak constraints on the SSA peak at $\sim90$ days post explosion.

Finally, we calculate a dimensionless normalisation of the wind density parameter $A_{\star} \propto n_{e}r^{2} \propto \dot{M}/v_{w}$ \citep{2000ApJ...536..195C},
\begin{equation}\label{eq:astar}
    A_{\star} = \frac{\dot{M}}{(5 \times 10^{11} ~{\rm g ~cm^{-1}})\times4\pi v_{w}} 
\end{equation}
where $A_{\star}=1$ for a Wolf-Rayet-like wind with $\dot{M}=10^{-5}$M$_{\odot}$yr$^{-1}$ and $v_{w}=1000$km\,s$^{-1}$. From our best fits to the radio data, we derive that AT\,2023fhn at $\sim$70--110 rest-frame days has $A_{\star}\sim1$ ($\sim0.1\times10^{-4}$~M$_{\odot}$ yr$^{-1}$ for $v_{\rm w}$ = 1000\,km~s$^{-1}$). This mass loss rate is consistent with that of Wolf-Rayet stars. As shown in Figure \ref{fig:CSM}, this density is also consistent with that of the other LFBOTs. The constraints on the synchrotron self-absorption peak at $\sim90$ days post explosion were unfortunately insufficient to constrain the density profile of the CSM around AT2023fhn.

\begin{table*}
	\centering
	\caption{Summary of radio SED fit results using the $\sim$90 day data, left, and $\sim$138\,day data, right, with equation \ref{eq:eq1}. Above the single solid lines we list the broken power-law fit parameters. Below, we list the inferred event properties at each epoch under the synchrotron blastwave model as described in Section \ref{sec:radio_results}, assuming equipartition ($\epsilon_{e}$=$\epsilon_{B}$=1/3). The uncertainties on the event properties (provided at 1$\sigma$) are statistical only, and are underestimated due to the presence of systematic errors arising from fixed values of $f$, $\epsilon_{\rm e}$ and $\epsilon_{\rm B}$. We also allowed $s$ to vary between 0 and 1, with $s=1$ providing the best-fit in each case. The (rest-frame) fit parameters $\nu_{\rm pk}$ and L$_{\rm \nu,pk}$ are defined at the intersection point of the two power laws \citep{1998ApJ...499..810C}. Below the double lines we give parameter values calculated with $\epsilon_{e}=0.1$ and $\epsilon_{B}=0.01$ (note that only the results with $\epsilon_{e}$=$\epsilon_{B}$=1/3 are plotted in the relevant figures).}
	\label{tab:results}
	\begin{tabular}{ccc} 
              & $t\sim90$\,days &  \\
		\hline
		\hline
		  Parameter & Unit & Value \\
		\hline
            $\nu_{\rm pk}$ & GHz & 3.6$\pm$0.2 \\
            $L_{\rm \nu,pk}$ & erg\,s$^{-1}$\,Hz$^{-1}$ & (3.6$\pm$0.3)$\times10^{29}$ \\
            $\beta_{1}$ & -- & 4$\pm$1 \\
            $\beta_{2}$ & -- & -0.6$\pm$0.1 \\
		\hline
            $R_{\rm p}$ & $10^{17}$\,cm & 1.3$\pm$0.1 \\
            $v/c = \Gamma \beta$ & -- & 0.72$\pm$0.03 \\
            $\dot{M}$/v$_{w}$ & 10$^{-4}$\,M$_{\odot}$\,yr$^{-1}$ / 1000\,km\,s$^{-1}$ & 0.08$\pm0.01$ \\
            $n_{e}$ & cm$^{-3}$ & 20$\pm$2 \\
            $B$ & G & 0.24$\pm0.02$ \\
            $U$ & $10^{49}$\,erg & 2.2$^{+0.3}_{-0.2}$ \\
		\hline
		\hline
            $R_{\rm p}$ & $10^{17}$\,cm & 1.2$\pm$0.1 \\
            $v/c = \Gamma \beta$ & -- & 0.63$\pm$0.03 \\
            $\dot{M}$/v$_{w}$ & 10$^{-4}$\,M$_{\odot}$\,yr$^{-1}$ / 1000\,km\,s$^{-1}$ & 0.94$\pm0.02$ \\
            $n_{e}$ & cm$^{-3}$ & (3.2$^{+0.4}_{-0.3}$)$\times10^{2}$ \\
            $B$ & G & 0.15$\pm0.01$ \\
            $U$ & $10^{49}$\,erg & 19$\pm$2 \\
		\hline
	\end{tabular}
        \quad
 	\begin{tabular}{ccc} 
                & $t\sim138$\,days &  \\
		\hline
		\hline
		  Parameter & Unit & Value \\
		\hline
            $\nu_{\rm pk}$ & GHz & 5.0$\pm$0.7 \\
            $L_{\rm \nu,pk}$ & erg\,s$^{-1}$\,Hz$^{-1}$ & (6.0$\pm$0.9)$\times10^{29}$ \\
            $\beta_{1}$ & -- & 2.8$\pm$0.8 \\
            $\beta_{2}$ & -- & -0.6$\pm$0.2 \\
		\hline
            $R_{\rm p}$ & $10^{17}$\,cm & 1.2$\pm$0.1 \\
            $v/c = \Gamma \beta$ & -- & 0.42$\pm$0.03 \\
            $\dot{M}$/v$_{w}$ & 10$^{-4}$\,M$_{\odot}$\,yr$^{-1}$ / 1000\,km\,s$^{-1}$ & 0.32$\pm$0.01 \\
            $n_{e}$ & cm$^{-3}$ & (1.0$^\pm$0.2)$\times10^{2}$ \\
            $B$ & G & 0.31$\pm$0.01 \\
            $U$ & $10^{49}$\,erg & 2.9$\pm$0.5 \\
		\hline
		\hline
            $R_{\rm p}$ & $10^{17}$\,cm & 1.1$\pm$0.1 \\
            $v/c = \Gamma \beta$ & -- & 0.37$\pm$0.03 \\
            $\dot{M}$/v$_{w}$ & 10$^{-4}$\,M$_{\odot}$\,yr$^{-1}$ / 1000\,km\,s$^{-1}$ & 4.0$\pm$0.1 \\
            $n_{e}$ & cm$^{-3}$ & (1.6$^{+0.3}_{-0.2}$)$\times10^{3}$ \\
            $B$ & G & 0.19$\pm$0.03 \\
            $U$ & $10^{49}$\,erg & 25$^{+5}_{-4}$ \\
		\hline
	\end{tabular}
\end{table*}

\section{Discussion}
In this Section, we discuss the host galaxy and derived properties of AT\,2023fhn in the context of the other LFBOTs discovered thus far. 

The host of AT\,2023fhn, taking the spiral and satellite as one interacting system, is broadly consistent with the host galaxies of core-collapse supernovae, and slightly above four other LFBOT hosts in terms of specific star formation rate (although below the host of ZTF\,18abvkwla). The host offset and faint, diffuse emission at the transient location (see Section \ref{sec:local}) are consistent with the tail of the core-collapse supernova distribution \citep[see also][]{2024MNRAS.527L..47C}. While the local and broader environment is consistent with a core-collapse origin, it is interesting to consider whether the high sSFR is related to tidal interaction between the spiral and satellite galaxy. Such interactions may be associated with an elevated tidal disrupted event (TDE) rate, which show a bias towards occurrence in the nuclei of post-starburst galaxies and galaxies undergoing interactions/mergers \citep{2016ApJ...818L..21F}. The non-nuclear location of AT\,2023fhn - at high offset from both the spiral and satellite - is difficult to explain in such a scenario \citep{2024MNRAS.527L..47C}, since galaxy-galaxy interactions are unlikely to have much impact on the TDE rate in massive clusters well away from the nuclei of either galaxy. On the other hand, galaxy scale interactions may increase the chance of finding a massive \citep[possibly IMBH-hosting,][]{2018ApJ...867..119F} cluster in such a location \citep[e.g.][]{2022A&A...657A..94N}, even if they do not increase the TDE rate {\it within} the cluster.

The measured optical magnitudes at the location of AT\,2023fhn (see Table \ref{tab:phot_table}) allow a contribution from a point source with absolute magnitude as bright as $\sim$-14. A significant contribution from a point source at the location of AT\,2023fhn is disfavoured (as the precise location has similar brightness to its immediate surroundings, as explained in Section \ref{sec:local}), but the presence of a globular cluster or ultra-compact dwarf galaxy - which may host massive black holes \citep[e.g.][]{2014Natur.513..398S} - cannot be ruled out. The presence of such an undetected cluster or ultra compact dwarf galaxy would be consistent with the upper limit on the black hole mass inferred from our late-time UV observations of $\lesssim 10^{5}$\,M$_{\odot}$ \citep{2024MNRAS.527.2452M}, given black hole - host galaxy/cluster mass relations \citep[][]{2013ARA&A..51..511K,2013A&A...555A..26L}.

The UV-optical, X-ray and radio evolution of AT\,2023fhn is broadly similar to other LFBOTs. Notably, however, the X-ray to UV luminosity ratio of AT\,2023fhn is an order of magnitude lower than AT\,2018cow at similar times, and up to 3 orders of magnitude lower than ZTF\,20acigmel. As we show in Section \ref{sec:radio_results}, it is difficult to attribute this variety to differences in the circumstellar medium density or blast-wave propagation, as AT\,2023fhn has a blast-wave velocity, energy and CSM comparable with other LFBOTs. This is consistent with the evidence from other LFBOTs thus far that the X-ray emission arises from a distinct mechanism, namely central engine activity. AT\,2023fhn is only the third LFBOT with a mildly relativistic outflow ($v=0.4^{+0.1}_{-0.2}c$), in common with CSS161010 and ZTF\,18abvkwla, demonstrating that the blast-wave is engine-driven. As can be seen in Figure \ref{fig:CSM}, all LFBOTs with sufficient constraints from radio observations thus far have a wind-like (in the sense that density decreases with distance), albeit not $r^{-2}$, circumstellar density profile. This suggests the CSM was produced by the progenitor system (i.e. through winds), rather than the explosion occurring in a pre-existing dense ISM, which would produce a flat density profile. 

An alternative explanation for the variety in UV/X-ray ratios is the viewing angle, where the asymmetric outflow and accretion disc are being viewed from different angles. In this interpretation, the viewing angle to AT\,2018cow was closer to perpendicular to the plane of the accretion disc \citep[although not exactly perpendicular,][]{2019ApJ...872...18M}. This conclusion was also reached by \citet{2023MNRAS.521.3323M} based on the high polarization \citep[and possibly for AT\,2022tsd, given the observation of late-time optical flares][]{2023Natur.623..927H}. AT\,2023fhn, meanwhile, would have been seen close to edge-on, well off-axis from any asymmetric outflow (e.g. from a choked jet) and with the inner accretion disc obscured \citep[where choked jets and/or the inner disc dominates the X-ray luminosity,][]{2020ApJ...889..166J}. The effect of viewing angle as an explanation for different LFBOT X-ray luminosities has also been posited by \citet{2022ApJ...932...84M} and \citet{2024ApJ...963L..24M}.

With the fundamental LFBOT requirement of a low $^{56}$Ni ejecta mass, and magnetar central engines struggling to explain all aspects of LFBOT phenomenology \citep[e.g. the late-time emission in AT2018cow,][]{2023ApJ...955...43C,2024ApJ...963L..13L}, constraints on the possible progenitor models are tightening. IMBH TDE models remain plausible, if a dense wind-like CSM can be produced (e.g. by the ejection of stripped mass during the disruption event). However, the star-forming nature of the host galaxy population, and the locations of LFBOTs within them, likely disfavour such an interpretation. Other plausible models include the delayed merger of black holes and Wolf-Rayet stars \citep{2022ApJ...932...84M}, and failed supernovae \citep{2019MNRAS.485L..83Q}. In AT\,2023fhn, the mass-loss wind parameter A$_{\star}\sim1$ - higher than many radio loud supernovae \citep{2006ApJ...651..381C} and collapsar GRBs \citep{2018ApJ...866..162G,2022MNRAS.515.2591C}. Such a dense circumstellar environment likely require a short-lived evolutionary stage with enhanced mass loss, for example pre-explosion winds from a blue supergiant or Wolf-Rayet star \citep{2019ApJ...872...18M}.

\section{Conclusions}
We list here our conclusions about the nature of AT\,2023fhn and its place in the context of other LFBOTs and extragalactic transients more generally,
   \begin{enumerate}
      \item Although relatively isolated compared with other LFBOTs and indeed most core-collapse supernovae, AT\,2023fhn is otherwise consistent with a core-collapse event, associated with a typical star-forming galaxy and located in a young (albeit diffuse) stellar population.
      \item The low X-ray to UV luminosity ratio demonstrates orders of magnitude of variety in this parameter among LFBOTs, which may be indicative of differences in viewing angle. In this interpretation, the relatively low X-ray luminosity of AT\,2023fhn is due to an edge-on viewing angle, such that the inner accretion disc is obscured and we are well off-axis from any choked jet/asymmetric outflow.
      \item The CSM properties are similar to previous LFBOTs, and are indicative of a dense surrounding medium. Given the wind-like n$_{\rm e}$ density profiles of other LFBOTs, and our n$_{\rm e}$ measurements of AT\,2023fhn which continue this trend, it is likely that the dense CSM was produced by wind-like mass-loss from the progenitor system itself (rather than the progenitor exploding in a pre-existing dense ISM).
      \item An IMBH TDE interpretation remains possible, only if there exists a pre-existing dense CSM or if the early stages of the tidal disruption produce such an environment. The host galaxy is likely undergoing tidal interactions, which may increase the chance of finding clusters (and hence TDEs) at high project offsets. While the non-nuclear location and host properties rate likely favour a core-collapse origin, the explosion of AT\,2023fhn in an undetected globular cluster or ultra-compact dwarf galaxy cannot be ruled out.
   \end{enumerate}
Despite mounting evidence, the origin of LFBOTs is still ambiguous. Two approaches will elucidate which of the proposed scenarios contribute to the LFBOT population. The first is to grow the sample, enabling statistically robust comparisons of offsets and host galaxy properties to be made with other classes of transient. This will be possible with the advent of new wide-field, deep sky surveys such as those performed by the Vera Rubin observatory \citep{2019ApJ...873..111I}. The second is to perform detailed studies of future local events - like AT2018cow - across the electromagnetic spectrum. Such events offer the best opportunity to search for underlying clusters, monitor the long-term evolution, understand the detailed emission physics, and ultimately, determine their progenitors. Although there is much progress still to be made, based on AT\,2023fhn and the growing population of LFBOTs, we deem a massive star progenitor with strong winds but low ejecta mass the most likely scenario. This favours models such as black hole/Wolf-Rayet mergers or failed supernovae.

\begin{acknowledgements}
      We thank the anonymous referee for their constructive feedback on this manuscript.\\
      
      A.A.C. acknowledges support from the European Space Agency (ESA) as an ESA Research Fellow. P.G.J.~has received funding from the European Research Council (ERC) under the European Union’s Horizon 2020 research and innovation programme (Grant agreement No.~101095973). P.J.G. is partly supported by NRF SARChI Grant 111692.\\

      Observations analysed in this work were taken by the NASA/ESA Hubble Space Telescope under program 17238. This research has made use of software provided by the Chandra X-ray Center (CXC) in the application of the CIAO package \citep{2006SPIE.6270E..1VF}. The National Radio Astronomy Observatory is a facility of the National Science Foundation operated under cooperative agreement by Associated Universities, Inc.\\
      
      Computing facilities were provided by the Scientific Computing Research Technology Platform of the University of Warwick. This research has made use of the Spanish Virtual Observatory (\url{https://svo.cab.inta-csic.es}) project funded by MCIN/AEI/10.13039/501100011033/ through grant PID2020-112949GB-I00. The Pan-STARRS1 Surveys (PS1) and the PS1 public science archive have been made possible through contributions by the Institute for Astronomy, the University of Hawaii, the Pan-STARRS Project Office, the Max-Planck Society and its participating institutes, the Max Planck Institute for Astronomy, Heidelberg and the Max Planck Institute for Extraterrestrial Physics, Garching, The Johns Hopkins University, Durham University, the University of Edinburgh, the Queen's University Belfast, the Harvard-Smithsonian Center for Astrophysics, the Las Cumbres Observatory Global Telescope Network Incorporated, the National Central University of Taiwan, the Space Telescope Science Institute, the National Aeronautics and Space Administration under Grant No. NNX08AR22G issued through the Planetary Science Division of the NASA Science Mission Directorate, the National Science Foundation Grant No. AST-1238877, the University of Maryland, Eotvos Lorand University (ELTE), the Los Alamos National Laboratory, and the Gordon and Betty Moore Foundation. This publication makes use of data products from the Wide-field Infrared Survey Explorer, which is a joint project of the University of California, Los Angeles, and the Jet Propulsion Laboratory/California Institute of Technology, funded by the National Aeronautics and Space Administration.

\end{acknowledgements}

\bibliographystyle{aa.bst}
\bibliography{2023fhn.bib}


\begin{appendix}

\section{SED fitting MCMC results}\label{app:A}
\begin{figure*}[t]
   \centering
   \includegraphics[width=\textwidth]{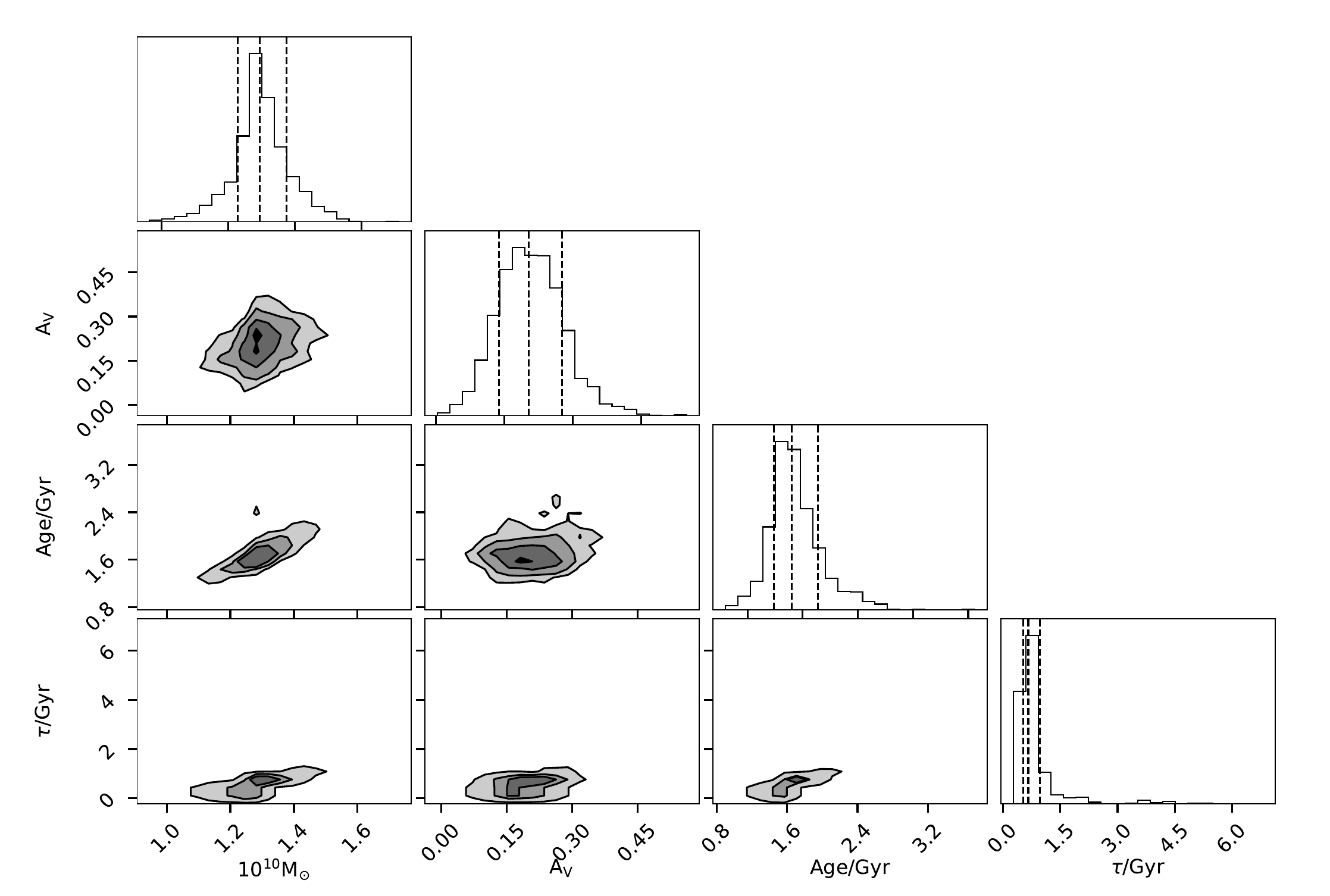}
      \caption{{\sc emcee} output from {\sc prospector} for the host of AT\,2023fhn. Produced using the {\sc corner.py} code \citep{corner} via {\sc prospector}. The metallicity is fixed at $0.5$Z$_{\odot}$.}
         \label{fig:posterior}
   \end{figure*}

In this appendix we provide the joint posterior parameter distributions for the host galaxy of AT\,2023fhn, in the form of a corner plot (Figure \ref{fig:posterior}) including stellar mass, metallicity, extinction, population age and timescale for an exponentially declining star formation history. These are provided as outputs from {\sc emcee} SED-fitting using {\sc prospector}. For the MCMC initial values (and flat priors) we use A$_{\rm V}=0.05$ (0<A$_{\rm V}$<2), t$_{\rm age}=1$\,Gyr and M$=10^{10}$\,M$_{\odot}$ ($10^{6}$<M/M$_{\odot}$<$10^{12}$), with a flat prior on $\tau$ of 0.1<$\tau$/Gyr<100. For our fiducial run, the redshift is fixed at $z=0.238$ and the metallicity at $Z=0.5Z_{\odot}$. Also provided in Table \ref{tab:hostresults_freeZ} are results when $Z$ is allowed to vary a free parameter.

\begin{table}
	\centering
	\caption{Host galaxy properties derived from {\sc prospector} SED fitting, as in Table \ref{tab:hostresults}, but allowing the metallicity $Z$ to vary.}
	\label{tab:hostresults_freeZ}
	\begin{tabular}{cc} 
		\hline
		\hline
		  Host property & Value \\
		\hline
            M$_{\star}$ / M$_{\odot}$ & (1.21$\pm$0.08) $\times 10^{10}$ \\
            SFR / M$_{\odot}$\,yr$^{-1}$  & 8.8$^{+1.3}_{-1.1}$ \\
            Z/Z$_{\odot}$  & 0.08$\pm0.02$ \\
            A$_{\rm V}$  & 0.47$^{+0.06}_{-0.06}$ \\
            t$_{\rm age}$/Gyr  & 1.9$^{+0.3}_{-0.4}$ \\
            $\tau$/Gyr & 1.1$^{+1.2}_{-0.6}$ \\
		\hline
    \end{tabular}
\end{table}

\end{appendix}

\end{document}